\documentclass[12pt,thmsa]{article}
\usepackage{amssymb}
\usepackage{sw20lart}

\input{tcilatex}
\input tcilatex

\makeatletter
 
\def\diagram{\m@th\leftwidth=\z@ \rightwidth=\z@ \topheight=\z@
\botheight=\z@ \setbox\@picbox\hbox\bgroup}
 
\def\enddiagram{\egroup\wd\@picbox\rightwidth\unitlength
\ht\@picbox\topheight\unitlength \dp\@picbox\botheight\unitlength
\hskip\leftwidth\unitlength\box\@picbox}
 
\def\bfig{\begin{diagram}}
\def\efig{\end{diagram}}
\newcount\wideness \newcount\leftwidth \newcount\rightwidth
\newcount\highness \newcount\topheight \newcount\botheight
 
\def\ratchet#1#2{\ifnum#1<#2 \global #1=#2 \fi}
 
\def\putbox(#1,#2)#3{%
\horsize{\wideness}{#3} \divide\wideness by 2
{\advance\wideness by #1 \ratchet{\rightwidth}{\wideness}}
{\advance\wideness by -#1 \ratchet{\leftwidth}{\wideness}}
\vertsize{\highness}{#3} \divide\highness by 2
{\advance\highness by #2 \ratchet{\topheight}{\highness}}
{\advance\highness by -#2 \ratchet{\botheight}{\highness}}
\put(#1,#2){\makebox(0,0){$#3$}}}
 
\def\putlbox(#1,#2)#3{%
\horsize{\wideness}{#3}
{\advance\wideness by #1 \ratchet{\rightwidth}{\wideness}}
{\ratchet{\leftwidth}{-#1}}
\vertsize{\highness}{#3} \divide\highness by 2
{\advance\highness by #2 \ratchet{\topheight}{\highness}}
{\advance\highness by -#2 \ratchet{\botheight}{\highness}}
\put(#1,#2){\makebox(0,0)[l]{$#3$}}}
 
\def\putrbox(#1,#2)#3{%
\horsize{\wideness}{#3}
{\ratchet{\rightwidth}{#1}}
{\advance\wideness by -#1 \ratchet{\leftwidth}{\wideness}}
\vertsize{\highness}{#3} \divide\highness by 2
{\advance\highness by #2 \ratchet{\topheight}{\highness}}
{\advance\highness by -#2 \ratchet{\botheight}{\highness}}
\put(#1,#2){\makebox(0,0)[r]{$#3$}}}

\def\adjust[#1]{} 
 
\newcount \coefa
\newcount \coefb
\newcount \coefc
\newcount\tempcounta
\newcount\tempcountb
\newcount\tempcountc
\newcount\tempcountd
\newcount\xext
\newcount\yext
\newcount\xoff
\newcount\yoff
\newcount\gap%
\newcount\arrowtypea
\newcount\arrowtypeb
\newcount\arrowtypec
\newcount\arrowtyped
\newcount\arrowtypee
\newcount\height
\newcount\width
\newcount\xpos
\newcount\ypos
\newcount\run
\newcount\rise
\newcount\arrowlength
\newcount\halflength
\newcount\arrowtype
\newdimen\tempdimen
\newdimen\xlen
\newdimen\ylen
\newsavebox{\tempboxa}%
\newsavebox{\tempboxb}%
\newsavebox{\tempboxc}%
 
\newdimen\w@dth
 
\def\setw@dth#1#2{\setbox\z@\hbox{\m@th$#1$}\w@dth=\wd\z@
\setbox\@ne\hbox{\m@th$#2$}\ifnum\w@dth<\wd\@ne \w@dth=\wd\@ne \fi
\advance\w@dth by 1.2em}
 
\def\t@^#1_#2{\allowbreak\def\n@one{#1}\def\n@two{#2}\mathrel
{\setw@dth{#1}{#2}
\mathop{\hbox to \w@dth{\rightarrowfill}}\limits
\ifx\n@one\empty\else ^{\box\z@}\fi
\ifx\n@two\empty\else _{\box\@ne}\fi}}
\def\t@@^#1{\@ifnextchar_{\t@^{#1}}{\t@^{#1}_{}}}
\def\to{\@ifnextchar^{\t@@}{\t@@^{}}}
 
\def\t@left^#1_#2{\def\n@one{#1}\def\n@two{#2}\mathrel{\setw@dth{#1}{#2}
\mathop{\hbox to \w@dth{\leftarrowfill}}\limits
\ifx\n@one\empty\else ^{\box\z@}\fi
\ifx\n@two\empty\else _{\box\@ne}\fi}}
\def\t@@left^#1{\@ifnextchar_{\t@left^{#1}}{\t@left^{#1}_{}}}
\def\toleft{\@ifnextchar^{\t@@left}{\t@@left^{}}}
 
\def\two@^#1_#2{\allowbreak
\def\n@one{#1}\def\n@two{#2}\mathrel{\setw@dth{#1}{#2}
\mathop{\vcenter{\lineskip\z@\baselineskip\z@
                 \hbox to \w@dth{\rightarrowfill}%
                 \hbox to \w@dth{\rightarrowfill}}%
       }\limits
\ifx\n@one\empty\else ^{\box\z@}\fi
\ifx\n@two\empty\else _{\box\@ne}\fi}}
\def\tw@@^#1{\@ifnextchar _{\two@^{#1}}{\two@^{#1}_{}}}
\def\two{\@ifnextchar ^{\tw@@}{\tw@@^{}}}
 
\def\tofr@^#1_#2{\def\n@one{#1}\def\n@two{#2}\mathrel{\setw@dth{#1}{#2}
\mathop{\vcenter{\hbox to \w@dth{\rightarrowfill}\kern-1.7ex
                 \hbox to \w@dth{\leftarrowfill}}%
       }\limits
\ifx\n@one\empty\else ^{\box\z@}\fi
\ifx\n@two\empty\else _{\box\@ne}\fi}}
\def\t@fr@^#1{\@ifnextchar_ {\tofr@^{#1}}{\tofr@^{#1}_{}}}
\def\tofro{\@ifnextchar^ {\t@fr@}{\t@fr@^{}}}

\def\mon{\mathop{\m@th\hbox to
      14.6\P@{\lasyb\char'51\hskip-2.1\P@$\arrext$\hss
$\mathord\rightarrow$}}\limits} 
\def\leftmono{\mathrel{\m@th\hbox to
14.6\P@{$\mathord\leftarrow$\hss$\arrext$\hskip-2.1\P@\lasyb\char'50%
}}\limits} 
\mathchardef\arrext="0200       

\def\settypes(#1,#2,#3){\arrowtypea#1 \arrowtypeb#2 \arrowtypec#3}
\def\settoheight#1#2{\setbox\@tempboxa\hbox{#2}#1\ht\@tempboxa\relax}%
\def\settodepth#1#2{\setbox\@tempboxa\hbox{#2}#1\dp\@tempboxa\relax}%
\def\settokens`#1`#2`#3`#4`{%
     \def\tokena{#1}\def\tokenb{#2}\def\tokenc{#3}\def\tokend{#4}}
\def\setsqparms[#1`#2`#3`#4;#5`#6]{%
\arrowtypea #1
\arrowtypeb #2
\arrowtypec #3
\arrowtyped #4
\width #5
\height #6
}
\def\setpos(#1,#2){\xpos=#1 \ypos#2}

\def\settriparms[#1`#2`#3;#4]{\settripairparms[#1`#2`#3`1`1;#4]}%
 
\def\settripairparms[#1`#2`#3`#4`#5;#6]{%
\arrowtypea #1
\arrowtypeb #2
\arrowtypec #3
\arrowtyped #4
\arrowtypee #5
\width #6
\height #6
}
 
\def\resetparms{\settripairparms[1`1`1`1`1;500]\width 500}
 
\resetparms
 
\def\mvector(#1,#2)#3{
\put(0,0){\vector(#1,#2){#3}}%
\put(0,0){\vector(#1,#2){26}}%
}
\def\evector(#1,#2)#3{{
\arrowlength #3
\put(0,0){\vector(#1,#2){\arrowlength}}%
\advance \arrowlength by-30
\put(0,0){\vector(#1,#2){\arrowlength}}%
}}
 
\def\horsize#1#2{%
\settowidth{\tempdimen}{$#2$}%
#1=\tempdimen
\divide #1 by\unitlength
}
 
\def\vertsize#1#2{%
\settoheight{\tempdimen}{$#2$}%
#1=\tempdimen
\settodepth{\tempdimen}{$#2$}%
\advance #1 by\tempdimen
\divide #1 by\unitlength
}
 
\def\putvector(#1,#2)(#3,#4)#5#6{{%
\ifnum3<\arrowtype
\putdashvector(#1,#2)(#3,#4)#5\arrowtype
\else
\ifnum\arrowtype<-3
\putdashvector(#1,#2)(#3,#4)#5\arrowtype
\else
\xpos=#1
\ypos=#2
\run=#3
\rise=#4
\arrowlength=#5
\ifnum \arrowtype<0
    \ifnum \run=0
        \advance \ypos by-\arrowlength
    \else
        \tempcounta \arrowlength
        \multiply \tempcounta by\rise
        \divide \tempcounta by\run
        \ifnum\run>0
            \advance \xpos by\arrowlength
            \advance \ypos by\tempcounta
        \else
            \advance \xpos by-\arrowlength
            \advance \ypos by-\tempcounta
        \fi
    \fi
    \multiply \arrowtype by-1
    \multiply \rise by-1
    \multiply \run by-1
\fi
\ifcase \arrowtype
\or \put(\xpos,\ypos){\vector(\run,\rise){\arrowlength}}%
\or \put(\xpos,\ypos){\mvector(\run,\rise)\arrowlength}%
\or \put(\xpos,\ypos){\evector(\run,\rise){\arrowlength}}%
\fi\fi\fi
}}
 
\def\putsplitvector(#1,#2)#3#4{
\xpos #1
\ypos #2
\arrowtype #4
\halflength #3
\arrowlength #3
\gap 140
\advance \halflength by-\gap
\divide \halflength by2
\ifnum\arrowtype>0
   \ifcase \arrowtype
   \or \put(\xpos,\ypos){\line(0,-1){\halflength}}%
       \advance\ypos by-\halflength
       \advance\ypos by-\gap
       \put(\xpos,\ypos){\vector(0,-1){\halflength}}%
   \or \put(\xpos,\ypos){\line(0,-1)\halflength}%
       \put(\xpos,\ypos){\vector(0,-1)3}%
       \advance\ypos by-\halflength
       \advance\ypos by-\gap
       \put(\xpos,\ypos){\vector(0,-1){\halflength}}%
   \or \put(\xpos,\ypos){\line(0,-1)\halflength}%
       \advance\ypos by-\halflength
       \advance\ypos by-\gap
       \put(\xpos,\ypos){\evector(0,-1){\halflength}}%
   \fi
\else \arrowtype=-\arrowtype
   \ifcase\arrowtype
   \or \advance \ypos by-\arrowlength
       \put(\xpos,\ypos){\line(0,1){\halflength}}%
       \advance\ypos by\halflength
       \advance\ypos by\gap
       \put(\xpos,\ypos){\vector(0,1){\halflength}}%
   \or \advance \ypos by-\arrowlength
       \put(\xpos,\ypos){\line(0,1)\halflength}%
       \put(\xpos,\ypos){\vector(0,1)3}%
       \advance\ypos by\halflength
       \advance\ypos by\gap
       \put(\xpos,\ypos){\vector(0,1){\halflength}}%
   \or \advance \ypos by-\arrowlength
       \put(\xpos,\ypos){\line(0,1)\halflength}%
       \advance\ypos by\halflength
       \advance\ypos by\gap
       \put(\xpos,\ypos){\evector(0,1){\halflength}}%
   \fi
\fi
}
 
\def\putmorphism(#1)(#2,#3)[#4`#5`#6]#7#8#9{{%
\run #2
\rise #3
\ifnum\rise=0
  \puthmorphism(#1)[#4`#5`#6]{#7}{#8}#9%
\else\ifnum\run=0
  \putvmorphism(#1)[#4`#5`#6]{#7}{#8}#9%
\else
\setpos(#1)%
\arrowlength #7
\arrowtype #8
\ifnum\run=0
\else\ifnum\rise=0
\else
\ifnum\run>0
    \coefa=1
\else
   \coefa=-1
\fi
\ifnum\arrowtype>0
   \coefb=0
   \coefc=-1
\else
   \coefb=\coefa
   \coefc=1
   \arrowtype=-\arrowtype
\fi
\width=2
\multiply \width by\run
\divide \width by\rise
\ifnum \width<0  \width=-\width\fi
\advance\width by60
\if l#9 \width=-\width\fi
\putbox(\xpos,\ypos){#4}
{\multiply \coefa by\arrowlength
\advance\xpos by\coefa
\multiply \coefa by\rise
\divide \coefa by\run
\advance \ypos by\coefa
\putbox(\xpos,\ypos){#5} }%
{\multiply \coefa by\arrowlength
\divide \coefa by2
\advance \xpos by\coefa
\advance \xpos by\width
\multiply \coefa by\rise
\divide \coefa by\run
\advance \ypos by\coefa
\if l#9%
   \putrbox(\xpos,\ypos){#6}%
\else\if r#9%
   \putlbox(\xpos,\ypos){#6}%
\fi\fi }%
{\multiply \rise by-\coefc
\multiply \run by-\coefc
\multiply \coefb by\arrowlength
\advance \xpos by\coefb
\multiply \coefb by\rise
\divide \coefb by\run
\advance \ypos by\coefb
\multiply \coefc by70
\advance \ypos by\coefc
\multiply \coefc by\run
\divide \coefc by\rise
\advance \xpos by\coefc
\multiply \coefa by140
\multiply \coefa by\run
\divide \coefa by\rise
\advance \arrowlength by\coefa
\ifcase\arrowtype
\or \put(\xpos,\ypos){\vector(\run,\rise){\arrowlength}}%
\or \put(\xpos,\ypos){\mvector(\run,\rise){\arrowlength}}%
\or \put(\xpos,\ypos){\evector(\run,\rise){\arrowlength}}%
\fi}\fi\fi\fi\fi}}

\newcount\numbdashes \newcount\lengthdash \newcount\increment
 
\def\howmanydashes{
\numbdashes=\arrowlength \lengthdash=40
\divide\numbdashes by \lengthdash
\lengthdash=\arrowlength
\divide\lengthdash by \numbdashes
\increment=\lengthdash
\multiply\lengthdash by 3
\divide\lengthdash by 5
}
 
\def\putdashvector(#1)(#2,#3)#4#5{%
\ifnum#3=0 \putdashhvector(#1){#4}#5
\else
\ifnum#2=0
\putdashvvector(#1){#4}#5\fi\fi}
 
\def\putdashhvector(#1,#2)#3#4{{%
\arrowlength=#3 \howmanydashes
\multiput(#1,#2)(\increment,0){\numbdashes}%
{\vrule height .4pt width \lengthdash\unitlength}
\arrowtype=#4 \xpos=#1
\ifnum\arrowtype<0 \advance\arrowtype by 7 \fi
\ifcase\arrowtype
\or \advance\xpos by 10
    \put(\xpos,#2){\vector(-1,0){\lengthdash}}
    \advance\xpos by 40
    \put(\xpos,#2){\vector(-1,0){\lengthdash}}
\or \advance \xpos by 10
    \put(\xpos,#2){\vector(-1,0){\lengthdash}}
    \advance\xpos by  \arrowlength
    \advance\xpos by  -50
    \put(\xpos,#2){\vector(-1,0){\lengthdash}}
\or \advance\xpos by 10
    \put(\xpos,#2){\vector(-1,0){\lengthdash}}
\or \advance\xpos by \arrowlength
    \advance\xpos by -\lengthdash
    \put(\xpos,#2){\vector(1,0){\lengthdash}}
\or {\advance\xpos by 10
    \put(\xpos,#2){\vector(1,0){\lengthdash}}}
    \advance\xpos by \arrowlength
    \advance\xpos by -\lengthdash
    \put(\xpos,#2){\vector(1,0){\lengthdash}}
\or \advance\xpos by \arrowlength
    \advance\xpos by -\lengthdash
    \put(\xpos,#2){\vector(1,0){\lengthdash}}
    \advance\xpos by -40
    \put(\xpos,#2){\vector(1,0){\lengthdash}}
   \fi
}}
 
\def\putdashvvector(#1,#2)#3#4{{%
\arrowlength=#3 \howmanydashes
\ypos=#2 \advance\ypos by -\arrowlength
\multiput(#1,#2)(0,\increment){\numbdashes}%
    {\vrule width .4pt height \lengthdash\unitlength}
\arrowtype=#4 \ypos=#2
\ifnum\arrowtype<0 \advance\arrowtype by 7 \fi
\ifcase\arrowtype
\or \advance\ypos by \arrowlength \advance\ypos by -40
    \put(#1,\ypos){\vector(0,1){\lengthdash}}
    \advance\ypos by -40
    \put(#1,\ypos){\vector(0,1){\lengthdash}}
\or \advance\ypos by 10
    \put(#1,\ypos){\vector(0,1){\lengthdash}}
    \advance\ypos by \arrowlength \advance\ypos by -40
    \put(#1,\ypos){\vector(0,1){\lengthdash}}
\or \advance\ypos by \arrowlength \advance\ypos by -40
    \put(#1,\ypos){\vector(0,1){\lengthdash}}
\or \advance\ypos by 10
    \put(#1,\ypos){\vector(0,-1){\lengthdash}}
\or \advance\ypos by 10
    \put(#1,\ypos){\vector(0,-1){\lengthdash}}
    \advance\ypos by \arrowlength \advance\ypos by -40
    \put(#1,\ypos){\vector(0,-1){\lengthdash}}
\or \advance\ypos by 10
    \put(#1,\ypos){\vector(0,-1){\lengthdash}}
    \advance\ypos by 40
    \put(#1,\ypos){\vector(0,-1){\lengthdash}}
\fi
}}
 
\def\puthmorphism(#1,#2)[#3`#4`#5]#6#7#8{{%
\xpos #1
\ypos #2
\width #6
\arrowlength #6
\arrowtype=#7
\putbox(\xpos,\ypos){#3\vphantom{#4}}%
{\advance \xpos by\arrowlength
\putbox(\xpos,\ypos){\vphantom{#3}#4}}%
\horsize{\tempcounta}{#3}%
\horsize{\tempcountb}{#4}%
\divide \tempcounta by2
\divide \tempcountb by2
\advance \tempcounta by30
\advance \tempcountb by30
\advance \xpos by\tempcounta
\advance \arrowlength by-\tempcounta
\advance \arrowlength by-\tempcountb
\putvector(\xpos,\ypos)(1,0)\arrowlength\arrowtype
\divide \arrowlength by2
\advance \xpos by\arrowlength
\vertsize{\tempcounta}{#5}%
\divide\tempcounta by2
\advance \tempcounta by20
\if a#8 %
   \advance \ypos by\tempcounta
   \putbox(\xpos,\ypos){#5}%
\else
   \advance \ypos by-\tempcounta
   \putbox(\xpos,\ypos){#5}%
\fi}}
 
\def\putvmorphism(#1,#2)[#3`#4`#5]#6#7#8{{%
\xpos #1
\ypos #2
\arrowlength #6
\arrowtype #7
\settowidth{\xlen}{$#5$}%
\putbox(\xpos,\ypos){#3}%
{\advance \ypos by-\arrowlength
\putbox(\xpos,\ypos){#4}}%
{\advance\arrowlength by-140
\advance \ypos by-70
\ifdim\xlen>0pt
   \if m#8%
      \putsplitvector(\xpos,\ypos)\arrowlength\arrowtype
   \else
   \putvector(\xpos,\ypos)(0,-1)\arrowlength\arrowtype
   \fi
\else
   \putvector(\xpos,\ypos)(0,-1)\arrowlength\arrowtype
\fi}%
\ifdim\xlen>0pt
   \divide \arrowlength by2
   \advance\ypos by-\arrowlength
   \if l#8%
      \advance \xpos by-40
      \putrbox(\xpos,\ypos){#5}%
   \else\if r#8%
      \advance \xpos by40
      \putlbox(\xpos,\ypos){#5}%
   \else
      \putbox(\xpos,\ypos){#5}%
   \fi\fi
\fi
}}
 
\def\putsquarep<#1>(#2)[#3;#4`#5`#6`#7]{{%
\setsqparms[#1]%
\setpos(#2)%
\settokens`#3`%
\puthmorphism(\xpos,\ypos)[\tokenc`\tokend`{#7}]{\width}{\arrowtyped}b%
\advance\ypos by \height
\puthmorphism(\xpos,\ypos)[\tokena`\tokenb`{#4}]{\width}{\arrowtypea}a%
\putvmorphism(\xpos,\ypos)[``{#5}]{\height}{\arrowtypeb}l%
\advance\xpos by \width
\putvmorphism(\xpos,\ypos)[``{#6}]{\height}{\arrowtypec}r%
}}
 
\def\putsquare{\@ifnextchar <{\putsquarep}{\putsquarep%
   <\arrowtypea`\arrowtypeb`\arrowtypec`\arrowtyped;\width`\height>}}
\def\square{\@ifnextchar< {\squarep}{\squarep
   <\arrowtypea`\arrowtypeb`\arrowtypec`\arrowtyped;\width`\height>}}
\def\squarep<#1>[#2`#3`#4`#5;#6`#7`#8`#9]{{
\setsqparms[#1]
\diagram
\putsquarep<\arrowtypea`\arrowtypeb`\arrowtypec`
\arrowtyped;\width`\height>
(0,0)[#2`#3`#4`{#5};#6`#7`#8`{#9}]
\enddiagram
}}                                                 
\def\putptrianglep<#1>(#2,#3)[#4`#5`#6;#7`#8`#9]{{%
\settriparms[#1]%
\xpos=#2 \ypos=#3
\advance\ypos by \height
\puthmorphism(\xpos,\ypos)[#4`#5`{#7}]{\height}{\arrowtypea}a%
\putvmorphism(\xpos,\ypos)[`#6`{#8}]{\height}{\arrowtypeb}l%
\advance\xpos by\height
\putmorphism(\xpos,\ypos)(-1,-1)[``{#9}]{\height}{\arrowtypec}r%
}}
 
\def\putptriangle{\@ifnextchar <{\putptrianglep}{\putptrianglep
   <\arrowtypea`\arrowtypeb`\arrowtypec;\height>}}
\def\ptriangle{\@ifnextchar <{\ptrianglep}{\ptrianglep
   <\arrowtypea`\arrowtypeb`\arrowtypec;\height>}}
\def\ptrianglep<#1>[#2`#3`#4;#5`#6`#7]{{
\settriparms[#1]
\diagram
\putptrianglep<\arrowtypea`\arrowtypeb`
\arrowtypec;\height>
(0,0)[#2`#3`#4;#5`#6`{#7}]
\enddiagram
}}                                            
 
\def\putqtrianglep<#1>(#2,#3)[#4`#5`#6;#7`#8`#9]{{%
\settriparms[#1]%
\xpos=#2 \ypos=#3
\advance\ypos by\height
\puthmorphism(\xpos,\ypos)[#4`#5`{#7}]{\height}{\arrowtypea}a%
\putmorphism(\xpos,\ypos)(1,-1)[``{#8}]{\height}{\arrowtypeb}l%
\advance\xpos by\height
\putvmorphism(\xpos,\ypos)[`#6`{#9}]{\height}{\arrowtypec}r%
}}
 
\def\putqtriangle{\@ifnextchar <{\putqtrianglep}{\putqtrianglep
   <\arrowtypea`\arrowtypeb`\arrowtypec;\height>}}
\def\qtriangle{\@ifnextchar <{\qtrianglep}{\qtrianglep
   <\arrowtypea`\arrowtypeb`\arrowtypec;\height>}}
\def\qtrianglep<#1>[#2`#3`#4;#5`#6`#7]{{
\settriparms[#1]
\width=\height                                
\diagram
\putqtrianglep<\arrowtypea`\arrowtypeb`
\arrowtypec;\height>
(0,0)[#2`#3`#4;#5`#6`{#7}]
\enddiagram
}}
 
\def\putdtrianglep<#1>(#2,#3)[#4`#5`#6;#7`#8`#9]{{%
\settriparms[#1]%
\xpos=#2 \ypos=#3
\puthmorphism(\xpos,\ypos)[#5`#6`{#9}]{\height}{\arrowtypec}b%
\advance\xpos by \height \advance\ypos by\height
\putmorphism(\xpos,\ypos)(-1,-1)[``{#7}]{\height}{\arrowtypea}l%
\putvmorphism(\xpos,\ypos)[#4``{#8}]{\height}{\arrowtypeb}r%
}}
 
\def\putdtriangle{\@ifnextchar <{\putdtrianglep}{\putdtrianglep
   <\arrowtypea`\arrowtypeb`\arrowtypec;\height>}}
\def\dtriangle{\@ifnextchar <{\dtrianglep}{\dtrianglep
   <\arrowtypea`\arrowtypeb`\arrowtypec;\height>}}
\def\dtrianglep<#1>[#2`#3`#4;#5`#6`#7]{{
\settriparms[#1]
\width=\height                                
\diagram
\putdtrianglep<\arrowtypea`\arrowtypeb`
\arrowtypec;\height>
(0,0)[#2`#3`#4;#5`#6`{#7}]
\enddiagram
}}
 
\def\putbtrianglep<#1>(#2,#3)[#4`#5`#6;#7`#8`#9]{{%
\settriparms[#1]%
\xpos=#2 \ypos=#3
\puthmorphism(\xpos,\ypos)[#5`#6`{#9}]{\height}{\arrowtypec}b%
\advance\ypos by\height
\putmorphism(\xpos,\ypos)(1,-1)[``{#8}]{\height}{\arrowtypeb}r%
\putvmorphism(\xpos,\ypos)[#4``{#7}]{\height}{\arrowtypea}l%
}}
 
\def\putbtriangle{\@ifnextchar <{\putbtrianglep}{\putbtrianglep
   <\arrowtypea`\arrowtypeb`\arrowtypec;\height>}}
\def\btriangle{\@ifnextchar <{\btrianglep}{\btrianglep
   <\arrowtypea`\arrowtypeb`\arrowtypec;\height>}}
\def\btrianglep<#1>[#2`#3`#4;#5`#6`#7]{{
\settriparms[#1]
\width=\height                               
\diagram
\putbtrianglep<\arrowtypea`\arrowtypeb`
\arrowtypec;\height>
(0,0)[#2`#3`#4;#5`#6`{#7}]
\enddiagram
}}
 
\def\putAtrianglep<#1>(#2,#3)[#4`#5`#6;#7`#8`#9]{{%
\settriparms[#1]%
\xpos=#2 \ypos=#3
{\multiply \height by2
\puthmorphism(\xpos,\ypos)[#5`#6`{#9}]{\height}{\arrowtypec}b}%
\advance\xpos by\height \advance\ypos by\height
\putmorphism(\xpos,\ypos)(-1,-1)[#4``{#7}]{\height}{\arrowtypea}l%
\putmorphism(\xpos,\ypos)(1,-1)[``{#8}]{\height}{\arrowtypeb}r%
}}
 
\def\putAtriangle{\@ifnextchar <{\putAtrianglep}{\putAtrianglep
   <\arrowtypea`\arrowtypeb`\arrowtypec;\height>}}
\def\Atriangle{\@ifnextchar <{\Atrianglep}{\Atrianglep
   <\arrowtypea`\arrowtypeb`\arrowtypec;\height>}}
\def\Atrianglep<#1>[#2`#3`#4;#5`#6`#7]{{
\settriparms[#1]
\width=\height                                     
\diagram
\putAtrianglep<\arrowtypea`\arrowtypeb`
\arrowtypec;\height>
(0,0)[#2`#3`#4;#5`#6`{#7}]
\enddiagram
}}
 
\def\putAtrianglepairp<#1>(#2)[#3;#4`#5`#6`#7`#8]{{%
\settripairparms[#1]%
\setpos(#2)%
\settokens`#3`%
\puthmorphism(\xpos,\ypos)[\tokenb`\tokenc`{#7}]{\height}{\arrowtyped}b%
\advance\xpos by\height
\puthmorphism(\xpos,\ypos)[\phantom{\tokenc}`\tokend`{#8}]%
{\height}{\arrowtypee}b%
\advance\ypos by\height
\putmorphism(\xpos,\ypos)(-1,-1)[\tokena``{#4}]{\height}{\arrowtypea}l%
\putvmorphism(\xpos,\ypos)[``{#5}]{\height}{\arrowtypeb}m%
\putmorphism(\xpos,\ypos)(1,-1)[``{#6}]{\height}{\arrowtypec}r%
}}
 
\def\putAtrianglepair{\@ifnextchar <{\putAtrianglepairp}{\putAtrianglepairp%
   <\arrowtypea`\arrowtypeb`\arrowtypec`\arrowtyped`\arrowtypee;\height>}}
\def\Atrianglepair{\@ifnextchar <{\Atrianglepairp}{\Atrianglepairp%
   <\arrowtypea`\arrowtypeb`\arrowtypec`\arrowtyped`\arrowtypee;\height>}}
 
\def\Atrianglepairp<#1>[#2;#3`#4`#5`#6`#7]{{
\settripairparms[#1]
\settokens`#2`
\width=\height                                
\diagram
\putAtrianglepairp                            
<\arrowtypea`\arrowtypeb`\arrowtypec`
\arrowtyped`\arrowtypee;\height>
(0,0)[{#2};#3`#4`#5`#6`{#7}]
\enddiagram
}}
 
\def\putVtrianglep<#1>(#2,#3)[#4`#5`#6;#7`#8`#9]{{%
\settriparms[#1]%
\xpos=#2 \ypos=#3
\advance\ypos by\height
{\multiply\height by2
\puthmorphism(\xpos,\ypos)[#4`#5`{#7}]{\height}{\arrowtypea}a}%
\putmorphism(\xpos,\ypos)(1,-1)[`#6`{#8}]{\height}{\arrowtypeb}l%
\advance\xpos by\height
\advance\xpos by\height
\putmorphism(\xpos,\ypos)(-1,-1)[``{#9}]{\height}{\arrowtypec}r%
}}
 
\def\putVtriangle{\@ifnextchar <{\putVtrianglep}{\putVtrianglep
   <\arrowtypea`\arrowtypeb`\arrowtypec;\height>}}
\def\Vtriangle{\@ifnextchar <{\Vtrianglep}{\Vtrianglep
   <\arrowtypea`\arrowtypeb`\arrowtypec;\height>}}
\def\Vtrianglep<#1>[#2`#3`#4;#5`#6`#7]{{
\settriparms[#1]
\width=\height                                 
\diagram
\putVtrianglep<\arrowtypea`\arrowtypeb`
\arrowtypec;\height>
(0,0)[#2`#3`#4;#5`#6`{#7}]
\enddiagram
}}
 
\def\putVtrianglepairp<#1>(#2)[#3;#4`#5`#6`#7`#8]{{
\settripairparms[#1]%
\setpos(#2)%
\settokens`#3`%
\advance\ypos by\height
\putmorphism(\xpos,\ypos)(1,-1)[`\tokend`{#6}]{\height}{\arrowtypec}l%
\puthmorphism(\xpos,\ypos)[\tokena`\tokenb`{#4}]{\height}{\arrowtypea}a%
\advance\xpos by\height
\puthmorphism(\xpos,\ypos)[\phantom{\tokenb}`\tokenc`{#5}]%
{\height}{\arrowtypeb}a%
\putvmorphism(\xpos,\ypos)[``{#7}]{\height}{\arrowtyped}m%
\advance\xpos by\height
\putmorphism(\xpos,\ypos)(-1,-1)[``{#8}]{\height}{\arrowtypee}r%
}}
 
\def\putVtrianglepair{\@ifnextchar <{\putVtrianglepairp}{\putVtrianglepairp%
    <\arrowtypea`\arrowtypeb`\arrowtypec`\arrowtyped`\arrowtypee;\height>}}
\def\Vtrianglepair{\@ifnextchar <{\Vtrianglepairp}{\Vtrianglepairp%
    <\arrowtypea`\arrowtypeb`\arrowtypec`\arrowtyped`\arrowtypee;\height>}}
\def\Vtrianglepairp<#1>[#2;#3`#4`#5`#6`#7]{{
\settripairparms[#1]
\settokens`#2`
\diagram
\putVtrianglepairp                             
<\arrowtypea`\arrowtypeb`\arrowtypec`
\arrowtyped`\arrowtypee;\height>
(0,0)[{#2};#3`#4`#5`#6`{#7}]
\enddiagram
}}

\def\putCtrianglep<#1>(#2,#3)[#4`#5`#6;#7`#8`#9]{{%
\settriparms[#1]%
\xpos=#2 \ypos=#3
\advance\ypos by\height
\putmorphism(\xpos,\ypos)(1,-1)[``{#9}]{\height}{\arrowtypec}l%
\advance\xpos by\height
\advance\ypos by\height
\putmorphism(\xpos,\ypos)(-1,-1)[#4`#5`{#7}]{\height}{\arrowtypea}l%
{\multiply\height by 2
\putvmorphism(\xpos,\ypos)[`#6`{#8}]{\height}{\arrowtypeb}r}%
}}
 
\def\putCtriangle{\@ifnextchar <{\putCtrianglep}{\putCtrianglep
    <\arrowtypea`\arrowtypeb`\arrowtypec;\height>}}
\def\Ctriangle{\@ifnextchar <{\Ctrianglep}{\Ctrianglep
    <\arrowtypea`\arrowtypeb`\arrowtypec;\height>}}
\def\Ctrianglep<#1>[#2`#3`#4;#5`#6`#7]{{
\settriparms[#1]
\width=\height                               
\diagram
\putCtrianglep<\arrowtypea`\arrowtypeb`
\arrowtypec;\height>
(0,0)[#2`#3`#4;#5`#6`{#7}]
\enddiagram
}}                                           
\def\putDtrianglep<#1>(#2,#3)[#4`#5`#6;#7`#8`#9]{{%
\settriparms[#1]%
\xpos=#2 \ypos=#3
\advance\xpos by\height \advance\ypos by\height
\putmorphism(\xpos,\ypos)(-1,-1)[``{#9}]{\height}{\arrowtypec}r%
\advance\xpos by-\height \advance\ypos by\height
\putmorphism(\xpos,\ypos)(1,-1)[`#5`{#8}]{\height}{\arrowtypeb}r%
{\multiply\height by 2
\putvmorphism(\xpos,\ypos)[#4`#6`{#7}]{\height}{\arrowtypea}l}%
}}
 
\def\putDtriangle{\@ifnextchar <{\putDtrianglep}{\putDtrianglep
    <\arrowtypea`\arrowtypeb`\arrowtypec;\height>}}
\def\Dtriangle{\@ifnextchar <{\Dtrianglep}{\Dtrianglep
   <\arrowtypea`\arrowtypeb`\arrowtypec;\height>}}
\def\Dtrianglep<#1>[#2`#3`#4;#5`#6`#7]{{
\settriparms[#1]
\width=\height                              
\diagram
\putDtrianglep<\arrowtypea`\arrowtypeb`
\arrowtypec;\height>
(0,0)[#2`#3`#4;#5`#6`{#7}]
\enddiagram
}}                                          
\def\setrecparms[#1`#2]{\width=#1 \height=#2}%
 
\def\recursep<#1`#2>[#3;#4`#5`#6`#7`#8]{{\m@th
\width=#1 \height=#2
\settokens`#3`
\settowidth{\tempdimen}{$\tokena$}
\ifdim\tempdimen=0pt
  \savebox{\tempboxa}{\hbox{$\tokenb$}}%
  \savebox{\tempboxb}{\hbox{$\tokend$}}%
  \savebox{\tempboxc}{\hbox{$#6$}}%
\else
  \savebox{\tempboxa}{\hbox{$\hbox{$\tokena$}\times\hbox{$\tokenb$}$}}%
  \savebox{\tempboxb}{\hbox{$\hbox{$\tokena$}\times\hbox{$\tokend$}$}}%
  \savebox{\tempboxc}{\hbox{$\hbox{$\tokena$}\times\hbox{$#6$}$}}%
\fi
\ypos=\height
\divide\ypos by 2
\xpos=\ypos
\advance\xpos by \width
\bfig
\putCtrianglep<-1`1`1;\ypos>(0,0)[`\tokenc`;#5`#6`{#7}]%
\puthmorphism(\ypos,0)[\tokend`\usebox{\tempboxb}`{#8}]{\width}{-1}b%
\puthmorphism(\ypos,\height)[\tokenb`\usebox{\tempboxa}`{#4}]{\width}{-1}a%
\advance\ypos by \width
\putvmorphism(\ypos,\height)[``\usebox{\tempboxc}]{\height}1r%
\efig
}}
 
\def\recurse{\@ifnextchar <{\recursep}{\recursep<\width`\height>}}
 
\def\puttwohmorphisms(#1,#2)[#3`#4;#5`#6]#7#8#9{{%
\puthmorphism(#1,#2)[#3`#4`]{#7}0a
\ypos=#2
\advance\ypos by 20
\puthmorphism(#1,\ypos)[\phantom{#3}`\phantom{#4}`#5]{#7}{#8}a
\advance\ypos by -40
\puthmorphism(#1,\ypos)[\phantom{#3}`\phantom{#4}`#6]{#7}{#9}b
}}
 
\def\puttwovmorphisms(#1,#2)[#3`#4;#5`#6]#7#8#9{{%
\putvmorphism(#1,#2)[#3`#4`]{#7}0a
\xpos=#1
\advance\xpos by -20
\putvmorphism(\xpos,#2)[\phantom{#3}`\phantom{#4}`#5]{#7}{#8}l
\advance\xpos by 40
\putvmorphism(\xpos,#2)[\phantom{#3}`\phantom{#4}`#6]{#7}{#9}r
}}
 
\def\puthcoequalizer(#1)[#2`#3`#4;#5`#6`#7]#8#9{{%
\setpos(#1)%
\puttwohmorphisms(\xpos,\ypos)[#2`#3;#5`#6]{#8}11%
\advance\xpos by #8
\puthmorphism(\xpos,\ypos)[\phantom{#3}`#4`#7]{#8}1{#9}
}}
 
\def\putvcoequalizer(#1)[#2`#3`#4;#5`#6`#7]#8#9{{%
\setpos(#1)%
\puttwovmorphisms(\xpos,\ypos)[#2`#3;#5`#6]{#8}11%
\advance\ypos by -#8
\putvmorphism(\xpos,\ypos)[\phantom{#3}`#4`#7]{#8}1{#9}
}}
 
\def\putthreehmorphisms(#1)[#2`#3;#4`#5`#6]#7(#8)#9{{%
\setpos(#1) \settypes(#8)
\if a#9 %
     \vertsize{\tempcounta}{#5}%
     \vertsize{\tempcountb}{#6}%
     \ifnum \tempcounta<\tempcountb \tempcounta=\tempcountb \fi
\else
     \vertsize{\tempcounta}{#4}%
     \vertsize{\tempcountb}{#5}%
     \ifnum \tempcounta<\tempcountb \tempcounta=\tempcountb \fi
\fi
\advance \tempcounta by 60
\puthmorphism(\xpos,\ypos)[#2`#3`#5]{#7}{\arrowtypeb}{#9}
\advance\ypos by \tempcounta
\puthmorphism(\xpos,\ypos)[\phantom{#2}`\phantom{#3}`#4]{#7}{\arrowtypea}{#9}
\advance\ypos by -\tempcounta \advance\ypos by -\tempcounta
\puthmorphism(\xpos,\ypos)[\phantom{#2}`\phantom{#3}`#6]{#7}{\arrowtypec}{#9}
}}
 
\def\setarrowtoks[#1`#2`#3`#4`#5`#6]{%
\def\toka{#1}
\def\tokb{#2}
\def\tokc{#3}
\def\tokd{#4}
\def\toke{#5}
\def\tokf{#6}
}
\def\hex{\@ifnextchar <{\hexp}{\hexp<1000`400>}}
\def\hexp<#1`#2>[#3`#4`#5`#6`#7`#8;#9]{%
\setarrowtoks[#9]
\yext=#2 \advance \yext by #2
\xext=#1 \advance\xext by \yext
\bfig
\putCtriangle<-1`0`1;#2>(0,0)[`#5`;\tokb``\tokd]
\xext=#1 \yext=#2 \advance \yext by #2
\putsquare<1`0`0`1;\xext`\yext>(#2,0)[#3`#4`#7`#8;\toka```\tokf]
\advance \xext by #2
\putDtriangle<0`1`-1;#2>(\xext,0)[`#6`;`\tokc`\toke]
\efig
}
\makeatother

\begin{document}

\title{The kinematical frame of Loop Quantum Gravity I}
\author{Andreas D\"{o}ring\linebreak \thanks{%
adoering@math.uni-frankfurt.de} , Hans F. de Groote\thanks{%
degroote@math.uni-frankfurt.de} \\
Fachbereich Mathematik,\\
Robert-Mayer-Stra\ss e 6-8,\\
Johann Wolfgang Goethe-Universit\"{a}t,\\
60054 Frankfurt am Main}
\date{26.12.2001 }
\maketitle

\begin{abstract}
In loop quantum gravity in the connection representation, the quantum
configuration space $\overline{\mathcal{A}/\mathcal{G}}$, which is a compact
space, is much larger than the classical configuration space $\mathcal{A}/%
\mathcal{G}$ of connections modulo gauge transformations. One finds that $%
\overline{\mathcal{A}/\mathcal{G}}$ is homeomorphic to the space $Hom(%
\mathcal{L}_{\ast },G))/Ad$. We give a new, natural proof of this result,
suggesting the extension of the hoop group $\mathcal{L}_{\ast }$ to a
larger, compact group $\mathcal{M}(\mathcal{L}_{\ast })$ that contains $%
\mathcal{L}_{\ast }$ as a dense subset. This construction is based on almost
periodic functions. We introduce the Hilbert algebra $L_{2}(\mathcal{M}(%
\mathcal{L}_{\ast }))$ of $\mathcal{M}(\mathcal{L}_{\ast })$ with respect to
the Haar measure $\xi $ on $\mathcal{M}(\mathcal{L}_{\ast })$. The measure $%
\xi $ is shown to be invariant under 3-diffeomorphisms. This is the first
step in a proof that $L_{2}(\mathcal{M}(\mathcal{L}_{\ast }))$ is the
appropriate Hilbert space for loop quantum gravity in the loop
representation. In a subsequent paper, we will reinforce this claim by
defining an extended loop transform and its inverse.
\end{abstract}

\pagebreak

\section{Introduction}

Over the last years, loop quantum gravity \cite{Rov98} has matured into a
powerful candidate for a theory of quantum gravity, with predictive power no
less than string theory, the other major approach to quantum gravity. There
are intriguing results concerning the nature of quantum geometry \cite%
{AshLew97a,AshLew97b} and black holes \cite{ABCK98,ABK00}. With the spin
network states \cite{Bae96a,Bae96b}, well defined orthonormal states of
quantum geometry are at hand, and spin foams seem to lead a way to a
covariant description of their evolution \cite{BarCra98,BarCra00,Bae98}.%
\newline

In this paper, we consider a much more humble aspect of loop quantum
gravity, namely the kinematical framework. Most of the rigorous work in loop
quantum gravity has been done in the connection representation \cite{ALMMT95}%
, but many developments were sparked by non-rigorous work in the loop
representation \cite{RovSmo93,RovSmo95}. So for example the seminal paper by
Rovelli and Smolin \cite{RovSmo90} - in which the loop representation was
defined - initiated a lot of research activity both in the loop and the
connection picture. As a result, a mathematically rigorous definition of the
kinematics of loop quantum gravity in the connection representation was
established, using $C^{\ast }$-algebra techniques \cite%
{AshIsh92,AshLew94,AshLew95,ALMMT95}.\newline

The quantum configuration space turned out to be a compact space containing
the classical configuration space $\mathcal{A}/\mathcal{G}$ of
Yang-Mills-theory and Ge-neral Relativity in the Ashtekar formulation as a
dense subspace. The Hilbert space of the unconstrained theory was found to
be some $L_{2}(\overline{\mathcal{A}/\mathcal{G}},d\mu )$, and the measure $%
\mu $ was explicitly constructed. This measure is invariant under
3-diffeomorphisms. The spin network states \cite{Bae96a,Bae96b} were found
to be an orthonormal basis of $L_{2}(\overline{\mathcal{A}/\mathcal{G}},d\mu
)$, and the diffeomorphism constraints were implemented \cite{ALMMT95},
although some questions seem to remain. Moreover, the so called loop
transform, also introduced in \cite{RovSmo90}, was rigorously defined \cite%
{AshLew94}. This transform translates from the connection to the loop
representation. So at least in the connection representation, the kinematics
of loop quantum gravity is well defined.\newline

On the other hand, in the loop representation much of the work does not
reach the same level of rigour. The states of loop quantum gravity in the
loop representation are defined rather symbolically \cite{RovSmo90,PieRov96}%
, as is the inner product of their Hilbert space \cite{Pie97}. There are
some ideas on how to define an inverse loop transform \cite{Thi95,Thi98}
that turn out to be rather involved technically.\newline

In this and a subsequent paper we will propose a slightly different view on
the kinematics of loop quantum gravity by making use of the compact group
associated to the hoop group $\mathcal{L}_{\ast }$ of holonomy equivalence
classes of piecewise analytic loops \cite{AshLew94}. One of the main results
in the connection picture is that the quantum configuration space $\overline{%
\mathcal{A}/\mathcal{G}}$ is homeomorphic to the space $Hom(\mathcal{L}%
_{\ast },G)/Ad$ of unitary equivalence classes of homomorphisms from the
hoop group $\mathcal{L}_{\ast }$ to the gauge group $G$, which is a closed
connected subgroup of some $U(N)$. (For loop quantum gravity, $G=SU(2)$). We
will make use of the fact that $\mathcal{L}_{\ast }$ can be embedded into a
larger, compact group $\mathcal{M}(\mathcal{L}_{\ast })$, containing $%
\mathcal{L}_{\ast }$ as a dense subgroup. $\mathcal{M}(\mathcal{L}_{\ast })$
is given as the spectrum of the $C^{\ast }$-algebra of almost periodic
functions on $\mathcal{L}_{\ast }$. It turns out that there is a
homeomorphism $\overline{\mathcal{A}/\mathcal{G}}\simeq Hom_{c}(\mathcal{M}(%
\mathcal{L}_{\ast }),G)/Ad$, too, where $Hom_{c}(\mathcal{M}(\mathcal{L}%
_{\ast }),G)/Ad$ is the space of equivalence classes of \emph{continuous }%
representations of $\mathcal{M}(\mathcal{L}_{\ast })$ by $G$-matrices.%
\newline

This minor change in the connection representation proves to be quite
useful. One considers the space $L_{2}(\mathcal{M}(\mathcal{L}_{\ast }))$,
the Hilbert algebra of $\mathcal{M}(\mathcal{L}_{\ast })$ with respect to
the Haar measure $\xi $ on $\mathcal{M}(\mathcal{L}_{\ast })$. The functions 
$t_{\rho }\in L_{2}(\mathcal{M}(\mathcal{L}_{\ast }))$, coming from the
Wilson loop functions $T_{\alpha }$ by switching the r\^{o}les of the
argument and the index, are shown to be orthonormal in an appropriate
normalization. Furthermore, the Haar measure $\xi $ turns out to be
invariant under 3-diffeomorphisms. This suggests to regard $L_{2}(\mathcal{M}%
(\mathcal{L}_{\ast }))$ as a candidate for the Hilbert space of loop quantum
gravity in the loop representation. In a subsequent paper, we will define an
extended loop transform from $L_{2}(\overline{\mathcal{A}/\mathcal{G}},d\mu
) $ to $L_{2}(\mathcal{M}(\mathcal{L}_{\ast }))$. The inverse transform is
given canonically. In the upcoming paper, we will also present some results
concerning the implementation of the diffeomorphism constraints, thus
defining the kinematical framework of loop quantum gravity rigorously in
both the loop and the connection representation.\newline

The plan of the paper is as follows: in section 2, we will consider almost
periodic functions and associated compact groups. In section 3, a new proof
is given for the fact that the quantum configuration space $\overline{%
\mathcal{A}/\mathcal{G}}$ of loop quantum gravity is homeomorphic to $Hom(%
\mathcal{L}_{\ast },G)/Ad$, the space of unitary equivalence classes of
homomorphisms (i.e. representations) from the hoop group $\mathcal{L}_{\ast
} $ to a closed connected subgroup $G$ of $U(N)$. It is also shown that $Hom(%
\mathcal{L}_{\ast },G)/Ad$ is canonically isomorphic to $Hom_{c}(\mathcal{M}(%
\mathcal{L}_{\ast }),G)/Ad$, the equivalence classes of continuous
representations of the compact group $\mathcal{M}(\mathcal{L}_{\ast })$
associated to the hoop group $\mathcal{L}_{\ast }$. So we get a homeomorphism%
\[
\overline{\mathcal{A}/\mathcal{G}}\longrightarrow Hom_{c}(\mathcal{M}(%
\mathcal{L}_{\ast }),G)/Ad, 
\]%
suggesting the enlargement of the hoop group to its associated compact group 
$\mathcal{M}(\mathcal{L}_{\ast })$. In section 4, we will define generalized
Wilson loop functions and introduce the Hilbert algebra $L_{2}(\mathcal{M}(%
\mathcal{L}_{\ast }))$ of $\mathcal{M}(\mathcal{L}_{\ast })$ with respect to
the normalized Haar measure $\xi $ on $\mathcal{M}(\mathcal{L}_{\ast })$.
The generalized Wilson loop functions are shown to be orthonormal (in an
appropriate normalization) and lie in the centre of this algebra. The Haar
measure on $L_{2}(\mathcal{M}(\mathcal{L}_{\ast }))$ is diffeomorphism
invariant. Section 5 gives a short outlook.

\section{Almost periodic functions and associated compact groups}

In this chapter we restate a proof known in the literature \cite{Loo53},
showing that to every topological group $H$ there exists a \textit{compact}
group $\mathcal{M}(H)$, characterized by a universal property, such that
there is a continuous homomorphism $\iota $ of $H$ onto a dense subset of $%
\mathcal{M}(H)$. $\mathcal{M}(H)$ is given as the spectrum of the $C^{\ast }$%
-algebra of almost periodic functions on $H$. Choosing $H=\mathcal{L}_{\ast
} $ in the following chapters, we will be able to shed some new light on the
kinematical structure of loop quantum gravity. The compact group $\mathcal{M}%
(H)$ associated to a topological group $H$ was mentioned in \cite{AshIsh92},
but not considered in the further development of the theory.\newline

For the convenience of the reader, we will first introduce almost periodic
functions:

\begin{definition}
Let $H$ be a group, $E$ a non-empty set and $f:H\rightarrow E$ a function.
For some fixed $s\in H$, let $f_{s}$ be the function on $H$ given by%
\[
f_{s}(x)=f(sx)\quad \forall x\in H. 
\]
$f_{s}$ is called the \textbf{left translate of }$f$\textbf{\ by }$s$.
\end{definition}

\begin{definition}
A \textbf{left almost periodic function} on a topological group $H$ is a
bounded function $f:H\rightarrow \mathbb{C}$ such that the set $%
S_{f}:=\{f_{s}\ |\ s\in H\}$ of left translates of $f$ is totally bounded
with respect to the supremum norm on $C_{b}(H)$, the space of bounded
continuous compex-valued functions on $H$.\newline
\end{definition}

Since $C_{b}(H)$ equipped with the supremum norm can be regarded as a metric
space, the total boundedness of $S_{f}$ is equivalent to the compactness of $%
\overline{S_{f}}\subseteq C_{b}(H)$. Let $\mathcal{F}(H)$ denote the set of
left almost periodic functions on $H$.\newline

One can define right almost periodic functions in an analogous manner. Both
definitions coincide, see \cite{HewRos63}, \S 18. From now on, we will just
speak of \textit{almost perodic functions}.\newline

Next we will state three propositions on almost periodic functions with the
aim of defining the compact group associated to a topological group. Proofs
can be found in \cite{Loo53}, ch. 41.

\begin{proposition}
$\mathcal{F}(H)\subseteq $ $C_{b}(H)$ is a commutative $C^{\ast }$-algebra
with unit element.
\end{proposition}

The Gelfand-Naimark theorem shows that to every commutative $C^{\ast }$%
-Algebra $\mathcal{F}(H)$ with unit element there exists a compact space $%
\mathcal{M}(H)$ such that\ $\mathcal{F}(H)$ is isometrically $\ast $%
-isomorphic to $C(\mathcal{M}(H))$. $\mathcal{M}(H)$ is the Gelfand spectrum
of $\mathcal{F}(H)$.

\begin{proposition}
There exists a continuous homomorphism $\iota $ from the group $H$ onto a
dense subspace of $\mathcal{M}(H)$.
\end{proposition}

One can show \cite{Loo53} that the group operation in $H$, regarded as a
subgroup of $\mathcal{M}(H)$, \ is \textit{uniformly }continuous in the
Gelfand topology on $\mathcal{M}(H)$ and hence can be extended to the whole
of $\mathcal{M}(H)$. In this way, $\mathcal{M}(H)$ aquires a group
structure. The central result is the following:\newline

\begin{proposition}
To every topological group $H$ there exists a compact group $\mathcal{M}(H)$
and a continuous homomorphism $\iota $ from $H$ onto a dense subgroup of $%
\mathcal{M}(H)$ such that for the pair $(\mathcal{M}(H),\iota )$ the
following universal condition holds: a function $f:H\rightarrow \mathbb{C}$
is almost periodic if and only if there exists a continuous function $g:%
\mathcal{M}(H)\rightarrow \mathbb{C}$ such that the following diagram is
commutative: 
\begin{center}
\settriparms[1`1`1;700]
\qtriangle[H`\QTR{cal}{M}(H)`\QTR{Bbb}{C};\iota`f`g]
\end{center}%
Since $\mathcal{M}(H)$ is compact, $g$ is determined by $f$ unambiguously.
We call $\mathcal{M}(H)$ \textbf{the compact group associated to }$H$.%
\newline
\end{proposition}

Let $G$ be a closed connected subgroup of some $U(N)$. When considering loop
quantum gravity, we will be interested in spaces $Hom(H,G)$ of homomorphisms
from a topological group $H$ (which will be chosen as $\mathcal{L}_{\ast }$,
the hoop group, or some subgroup of it) to the compact group $G$. For these
spaces $Hom(H,G)$ of representations we have

\begin{lemma}
Let $\mathcal{M}(H)$ be the compact group associated to $H$. Then there
exists a canonical bijection between $Hom(H,G)$ and $Hom_{c}(\mathcal{M}%
(H),G)$, the set of all continuous homomorphisms $\mathcal{M}(H)\rightarrow
G $.
\end{lemma}

\begin{proof}
Let $\varphi \in Hom(H,G)$. Then there exists a unique continuous
homomorphism%
\[
\mathcal{M}(\varphi ):\mathcal{M}(H)\rightarrow G
\]%
making the diagram

\begin{center}
\settriparms[1`-1`-1;700]
\ptriangle[\QTR{cal}{M}(H)`G`H;\QTR{cal}{M}(\varphi )`\iota`\varphi]
\end{center}%

commutative. This can be seen as follows: since $G$ is a closed subgroup of
some $U(N)$, \ one can consider the matrix elements%
\[
\rho _{ij}(x):=\left\langle \rho (x)e_{i},e_{j}\right\rangle ,
\]%
where $\{e_{i}\}$ is the canonical basis of the space $\mathbb{C}^{N}$ the
group $G$ acts on. Let $s\in H$. One has a left translate%
\[
\rho _{ij}^{s}(x)=\rho _{ij}(sx)=\left\langle \rho (s)\rho
(x)e_{i},e_{j}\right\rangle .
\]%
We want to show that the $\rho _{ij}$ are almost periodic, so regard%
\begin{eqnarray*}
|\rho _{ij}^{s}(x)-\rho _{ij}^{t}(x)| &=&|\left\langle \rho (s)\rho
(x)e_{i},e_{j}\right\rangle -\left\langle \rho (t)\rho
(s)e_{i},e_{j}\right\rangle | \\
&=&|\left\langle (\rho (s)-\rho (t))\rho (x)e_{i},e_{j}\right\rangle | \\
&\leq &|(\rho (s)-\rho (t))\rho (x)e_{i}| \\
&\leq &|\rho (s)-\rho (t)| \\
&=&|\rho (s)(I_{N}-\rho (s^{-1}t))| \\
&\leq &|I_{N}-\rho (s^{-1}t)|.
\end{eqnarray*}%
It follows that%
\[
|\rho _{ij}^{s}-\rho _{ij}^{t}|_{\infty }\leq |I_{N}-\rho (s^{-1}t)|.
\]%
Let $\varepsilon >0$ and%
\[
U_{0}:=\{g\in G|\ |I_{N}-g|<\varepsilon \}.
\]%
Since $G$ is compact, there exist finitely many $x_{1},...,x_{n}\in H$ such
that%
\[
\rho (H)\subseteq \bigcup_{k=1}^{n}\rho (x_{i})U_{0}.
\]%
This means that for any $x\in H$ there exists some $k\leq n$ such that%
\begin{eqnarray*}
\rho (x) &\in &\rho (x_{k})U_{0} \\
\Leftrightarrow \rho (x_{k}^{-1}x) &\in &U_{0},
\end{eqnarray*}%
so we have%
\[
|\rho _{ij}^{x_{k}}-\rho _{ij}^{x}|_{\infty }\leq |I_{N}-\rho
(x_{k}^{-1}x)|<\varepsilon ,
\]%
i.e., $\rho _{ij}$ is almost periodic. It follows that $\rho _{ij}$ can be
extended to a continuous function $\mathcal{M}(\rho _{ij}):\mathcal{M}%
(H)\rightarrow \mathbb{C}$ unambiguously. Of course, one has $\mathcal{M}%
(\rho _{ij})\circ \iota =\rho _{ij}$. Define 
\[
\mathcal{M}(\rho )(x)_{ij}:=\mathcal{M(}\rho _{ij})(x)\quad \forall x\in
\iota (H).
\]%
Since $\iota (H)$ is dense in $\mathcal{M}(H)$, the mapping $\mathcal{M}%
(\rho ):\mathcal{M}(H)\rightarrow G$ is a continuous representation. This
gives a bijection $\varphi \rightarrow \mathcal{M}(\varphi )$ between $%
Hom(H,G)$ and $Hom_{c}(\mathcal{M}(H),G)$. Since $\mathcal{M}(\varphi )$ is
defined by a universal property, this bijection is canonical.
\end{proof}

\begin{remark}
The bijection $\varphi \rightarrow \mathcal{M}(\varphi )$ is natural in the
category theoretical sense.
\end{remark}

\begin{proof}
Let $\phi :\Gamma _{1}\rightarrow \Gamma _{2}$ be a continuous homomorphism
of topological groups. Since $Hom(\_,G)$ is a cofunctor, one has a natural
map%
\[
\phi _{\ast }:Hom(\Gamma _{2},G)\rightarrow Hom(\Gamma _{1},G),\quad \psi
\mapsto \psi \circ \phi .
\]%
The universal property of associated compact groups renders the following
diagram commutative:

\begin{center}
\xext=2100 \yext=500
\xoff=0 \yoff=0
\begin{picture}(\xext,\yext)(\xoff,\yoff)
\resetparms
\putmorphism(418,0)(0,1)[`\QTR{cal}{M}(\Gamma _{1})`\iota _{1}]{450}1l
\putmorphism(800,-300)(4,1)[``\QTR{cal}{M}(\psi \circ \phi )]{720}1r
\setsqparms[1`-1`-1`1;900`400]
\square[\QTR{cal}{M}(\Gamma _{1})`\QTR{cal}{M}(\Gamma _{2})`\Gamma _{1}`\Gamma _{2};\QTR{cal}{M}(\iota _{2}\circ \phi )`\iota _{1}`\iota _{2}`\phi ]
\putmorphism(-120,0)(1,0)[`G`\psi ]{500}1a
\putmorphism(-30,450)(1,-1)[``\QTR{cal}{M}(\psi )]{450}1r
\end{picture}
\end{center}%
\vspace{1.8cm}

Uniqueness gives%
\[
\mathcal{M}(\psi \circ \phi )=\mathcal{M}(\psi )\circ \mathcal{M}(\iota
_{2}\circ \phi ),
\]%
which means that the following diagram is commutative:

\begin{center}
\setsqparms[1`1`1`1;1300`700]
\square[Hom(\Gamma _{2},G)`Hom(\Gamma _{1},G)`Hom_{c}(\QTR{cal}{M}(\Gamma _{2}),G)`Hom_{c}(\QTR{cal}{M}(\Gamma _{1}),G);\phi _{\ast }`\QTR{cal}{M}`\QTR{cal}{M}`\QTR{cal}{M}(\iota _{2}\circ \phi )_{\ast }]
\end{center}%
\end{proof}

\begin{remark}
Let $Hom(\mathcal{M}(H),G)$ be the set of \emph{all} homomorphisms $\mathcal{%
M}(H)\rightarrow G$, $\varphi \in Hom(H,G)$. Then the set%
\[
\{\psi \in Hom(\mathcal{M}(H),G)\ |\ \psi \circ \iota =\varphi \} 
\]%
contains exactly one continuous homomorphism, which is $\mathcal{M}(\varphi
) $.\newline
\end{remark}

\section{The homeomorphism between $Spec(C^{\ast }(\mathcal{HA}))$ and $Hom(%
\mathcal{L}_{\ast },G)/Ad$}

In this section we give a new, natural proof showing the homeomorphism
between the spectrum $\overline{\mathcal{A}/\mathcal{G}}:=Spec(C^{\ast }(%
\mathcal{HA}))$ of the Ashtekar-Isham algebra $C^{\ast }(\mathcal{HA})$ (for
definitions see below) and the space $Hom(\mathcal{L}_{\ast },G)/Ad$ of
unitary equivalence classes of homomorphisms from the hoop group $\mathcal{L}%
_{\ast }$ to a closed connected subgroup $G$ of $U(N)$. The space $\overline{%
\mathcal{A}/\mathcal{G}}$ of generalized connections modulo generalized
gauge transformations serves as the quantum configuration space in loop
quantum gravity \cite{AshLew94}. $\overline{\mathcal{A}/\mathcal{G}}$
contains the classical configuration space $\mathcal{A}/\mathcal{G}$ of
Yang-Mills theory and General Relativity in the Ashtekar formulation as a
dense subset. Every element of $\mathcal{A}/\mathcal{G}$ induces an element
of $Hom(\mathcal{L}_{\ast },G)/Ad$ in a natural way: consider the mapping%
\begin{eqnarray*}
\mathcal{A}/\mathcal{G} &\longrightarrow &Hom(\mathcal{L}_{\ast },G)/Ad, \\
\lbrack A]_{\mathcal{G}} &\longmapsto &\frac{1}{N}tr(\rho (H(\_,A))),
\end{eqnarray*}%
where $A$ is some representative of $[A]_{\mathcal{G}}\in \mathcal{A}/%
\mathcal{G}$,$\ H(\_,A):\mathcal{L}_{\ast }\rightarrow G$ is the holonomy
mapping with respect to the connection $A$ and $\rho :G\rightarrow M(\mathbb{%
C},N)$ is some representation of $G$ by complex $(N\times N)$ matrices. The
mapping is well defined since the trace of the holonomy is gauge invariant.
The idea now is to show that every element of $\overline{\mathcal{A}/%
\mathcal{G}}$, not just $\mathcal{A}/\mathcal{G}$, can be seen as the (trace
of) some homomorphism from the hoop group $\mathcal{L}_{\ast }$ to $G$. This
can be done by explicitly constructing the homomorphism belonging to each $%
\overline{A}\in \overline{\mathcal{A}/\mathcal{G}}$ as in \cite{AshIsh92} or
by using projective techniques \cite{AshLew95}. In these proofs, one has to
consider Mandelstam identities, the reconstruction results by Giles \cite%
{Gil81} and finite dimensional (or finitely generated) objects like tame
subgroups of $\mathcal{L}_{\ast }$ and cylinder functions on them which give
the interesting objects in the projective limit. A recent paper along these
lines, extending some of the known results, is \cite{AbbMan01}. While these
techniques prove to be extremely useful for the further development of the
theory, they are not essential to the fact that $\overline{\mathcal{A}/%
\mathcal{G}}$ is homeomorphic to $Hom(\mathcal{L}_{\ast },G)/Ad$, as will be
shown below. In a sense, our proof is more natural than the existing ones,
since it works directly with the interesting objects $\overline{\mathcal{A}/%
\mathcal{G}}$ and $Hom(\mathcal{L}_{\ast },G)/Ad$.\newline

But a new proof for an old result is not so exiting, in general, as long as
it does not give some new insight. Now it turns out that at some
intermediate step in our proof it is necessary to refer to the compact group 
$\mathcal{M}(\mathcal{L}_{\ast })$ associated to the hoop group $\mathcal{L}%
_{\ast }$. While this may seem a technicality at first, it allows a slight
change of perspective when regarding the quantum configuration space $%
\overline{\mathcal{A}/\mathcal{G}}$: since there is a canonical bijection
between $Hom(\mathcal{L}_{\ast },G)/Ad$ and $Hom_{c}(\mathcal{M}(\mathcal{L}%
_{\ast }),G)/Ad$, the unitary equivalence classes of \textit{continuous}
homomorphisms from $\mathcal{M}(\mathcal{L}_{\ast })$ to $G$, $\overline{%
\mathcal{A}/\mathcal{G}}$ is homoemorphic to $Hom_{c}(\mathcal{M}(\mathcal{L}%
_{\ast }),G)/Ad$, too. This fact will be used in the following section.%
\newline

Let $G$ be a Hausdorff topological group, $\Gamma $ a group. By $G^{\Gamma }$
we denote the space of all maps from $\Gamma $ to $G$.

\begin{theorem}
Let $\Gamma $ be a group, $G$ a Hausdorff topological group. Then the space $%
Hom(\Gamma ,G)\subseteq G^{\Gamma }$ is closed in the product topology. It
follows that $Hom(\Gamma ,G)$ is compact if $G$ is compact.
\end{theorem}

\begin{proof}
Let $H\in \overline{Hom(\Gamma ,G)}$. $H$ is a map $\Gamma \rightarrow G$,
and one has to show that 
\[
H(\alpha \beta )=H(\alpha )H(\beta )
\]%
for all $\alpha ,\beta \in \Gamma $. Let $U\subseteq G$ be an open
neighbourhood of the neutral element $e$ of $G$, $V$ a symmetric open
neighbourhood of $e$ such that $V^{3}\subseteq U$ and $W\subseteq V$ an
open neighbourhood of $e$ with $WH(\beta )\subseteq H(\beta )V$. The
existence of $W$ follows from the continuity of the multiplication in $G$.
Let $pr_{\alpha }:G^{\Gamma }\rightarrow G$ be the projection $f\rightarrow
f(\alpha )$ to the coordinate $\alpha $ of $f\in G^{\Gamma }$. Then 
\[
\mathcal{O}:=pr_{\alpha }^{-1}(H(\alpha )W)\cap pr_{\beta }^{-1}(H(\beta
)W)\cap pr_{\alpha \beta }^{-1}(H(\alpha \beta )W)
\]%
is an open neighbourhood of $H$ in the product topology. Let $\varphi \in
Hom(\Gamma ,G)\cap \mathcal{O}$. Then 
\begin{eqnarray*}
\varphi (\alpha ) &\in &H(\alpha )W, \\
\varphi (\beta ) &\in &H(\beta )W, \\
\varphi (\alpha \beta ) &\in &H(\alpha \beta )W,
\end{eqnarray*}%
that is, there exist $g_{\alpha },g_{\beta },g_{\alpha \beta }\in W$ such
that 
\begin{eqnarray*}
\varphi (\alpha ) &=&H(\alpha )g_{\alpha }, \\
\varphi (\beta ) &=&H(\beta )g_{\beta }, \\
\varphi (\alpha \beta ) &=&H(\alpha \beta )g_{\alpha \beta },
\end{eqnarray*}%
and it follows that 
\begin{eqnarray*}
H(\alpha \beta ) &=&\varphi (\alpha \beta )g_{\alpha \beta }^{-1} \\
&=&\varphi (\alpha )\varphi (\beta )g_{\alpha \beta }^{-1} \\
&=&H(\alpha )g_{\alpha }H(\beta )g_{\beta }g_{\alpha \beta }^{-1} \\
&\in &H(\alpha )H(\beta )V\cdot V\cdot V \\
&\subseteq &H(\alpha )H(\beta )U,
\end{eqnarray*}%
where we have used $WH(\beta )\subseteq H(\beta )V$ in the fourth step.
Since $U$ is an arbitrary neighbourhood of $e$, one finds that 
\[
H(\alpha \beta )=H(\alpha )H(\beta ).
\]
\end{proof}

\begin{lemma}
Let $\mathcal{A},\mathcal{B}$ be algebras over a field $F$. Let $\mathcal{E}$
be a linear system of generators for $\mathcal{A}$, that is 
\[
\mathcal{A}=lin\prod\nolimits_{\mathcal{E}}\ , 
\]%
where $\prod\nolimits_{\mathcal{E}}$ is the set of finite products of
elements of $\mathcal{E}$. We further assume that $\mathcal{A}$ has a
neutral element $1$ and that $1\in \mathcal{E}$. The map 
\[
\varphi :\mathcal{E}\rightarrow \mathcal{B} 
\]%
can be extended to an algebra homomorphism 
\[
\overline{\varphi }:\mathcal{A}\rightarrow \mathcal{B}, 
\]%
if and only if the following conditions hold:

\begin{enumerate}
\item For $a_{1},...,a_{m},b_{1},...,b_{n}\in \mathcal{E}$ with $%
a_{1}...a_{m}=b_{1}...b_{n}$ one has $\varphi (a_{1})...\varphi
(a_{m})=\varphi (b_{1})...\varphi (b_{n})$.

\item Let $\sum_{k}c_{k}p_{k}=0$ for some $p_{k}\in \prod\nolimits_{\mathcal{%
E}}$, $p_{k}=a_{k_{1}}...a_{k_{n_{k}}}$, $c_{k}\in K$. Then $%
\sum_{k}c_{k}\varphi (a_{k_{1}})...\varphi (a_{k_{n_{k}}})=0$.
\end{enumerate}
\end{lemma}

\begin{proof}
If there is a homomorphic extension $\overline{\varphi }:\mathcal{A}%
\rightarrow \mathcal{B}$ of $\varphi $, the conditions (1) and (2) have to
hold. On the other hand, assume that (1) and (2) are fulfilled for some $%
\varphi :\mathcal{E}\rightarrow \mathcal{B}$. Because of condition (1), $%
\varphi $ can be extended to a map 
\[
\varphi :\prod\nolimits_{\mathcal{E}}\rightarrow \mathcal{B}
\]%
by setting 
\[
\varphi (a_{1}...a_{m}):=\varphi (a_{1})...\varphi (a_{m}).
\]%
$\varphi $ is defined on a linear system of generators $\prod\nolimits_{%
\mathcal{E}}$ of $\mathcal{A}$ and has the property 
\[
\sum\nolimits_{k}c_{k}p_{k}=0\Rightarrow \sum\nolimits_{k}c_{k}\varphi
(p_{k})=0
\]%
because of condition (2). This condition assures that $\varphi $ can be
extended unambigously to a linear map $\overline{\varphi }:\mathcal{A}%
\rightarrow \mathcal{B}$. In order to see this, consider the vector space $%
\mathcal{F}(\Pi _{\mathcal{E}})$ that is freely generated by $\Pi _{\mathcal{%
E}}$. A concrete model of $\mathcal{F}(\Pi _{\mathcal{E}})$ is the space of
all functions $f:\Pi _{\mathcal{E}}\rightarrow \mathbb{C}$ of finite
support. Any function of this type can be written as $f=\sum_{p\in \Pi _{%
\mathcal{E}}}c_{p}\chi _{\{p\}}$ with $c_{p}=0$ for almost all $p\in \Pi _{%
\mathcal{E}}$, where $\chi _{\{p\}}$ denotes the characteristic function of $%
\{p\}$: $\chi _{\{p\}}(q)=\delta _{pq}$. We have a canonical surjective
linear mapping 
\[
e:\mathcal{F}(\Pi _{\mathcal{E}})\rightarrow \mathcal{A}
\]%
defined by 
\[
e(\sum_{p\in \Pi _{\mathcal{E}}}c_{p}\chi _{\{p\}}):=\sum_{p\in \Pi _{%
\mathcal{E}}}c_{p}p.
\]%
Therefore 
\[
\mathcal{A}\simeq \mathcal{F}(\Pi _{\mathcal{E}})/ker\ e.
\]%
Define $\varphi _{\mathcal{F}}:\mathcal{F}(\Pi _{\mathcal{E}})\rightarrow 
\mathcal{B}$ by%
\[
\varphi _{\mathcal{F}}(\sum_{p}c_{p}\chi _{\{p\}}):=\sum_{p}c_{p}\varphi (p).
\]%
The condition on $\varphi $ implies that $\varphi _{\mathcal{F}}$ induces a
linear mapping 
\[
\tilde{\varphi}:\mathcal{F}(\Pi _{\mathcal{E}})/ker\ e\rightarrow \mathcal{B}
\]%
and, by the isomorphism above, a linear map 
\[
\bar{\varphi}:\mathcal{A}\rightarrow \mathcal{B}
\]%
such that $\bar{\varphi}(a_{1}\cdots a_{m})=\varphi (a_{1})\cdots \varphi
(a_{m})$ for all $a_{1},\ldots ,a_{m}\in \mathcal{E}$. Obviously, $\bar{%
\varphi}$ is also multiplicative: 
\[
\bar{\varphi}(ab)=\bar{\varphi}(a)\bar{\varphi}(b)
\]%
for all $a,b\in \mathcal{A}$.
\end{proof}\newline

From now on, let $G$ be a closed connected subgroup of some unitary group $%
U(N)$. The following result by Ashtekar and Lewandowski \cite{AshLew94} is
central to all our further considerations:

\begin{theorem}
(Interpolation theorem, Ashtekar and Lewandowski.) Let $\mathcal{L}_{\ast }$
be the hoop group of \emph{piecewise analytic} hoops in the manifold $M$
with base point $\ast $, $P(M,G)$ a principal bundle over $M$ with structure
group $G$. Let $\varphi :\mathcal{L}_{\ast }\rightarrow G$ be a group
homomorphism. For every finite subset $\{\alpha _{1},...,\alpha _{n}\}$ of $%
\mathcal{L}_{\ast }$, there exists a connection $A$ in $P(M,G)$, such that%
\[
\varphi (\alpha _{i})=H(\alpha _{i},A) 
\]%
for all $i=1,...,n$.\newline
\end{theorem}

\begin{remark}
There is another version of this theorem: let $x$ denote the constant hoop.
Given a finite subset $\{\alpha _{1},...,\alpha _{n}\}$ of $\mathcal{L}%
_{\ast }$, $\alpha _{i}\neq x$, and an $n$-tuple $\{g_{1},...,g_{n}\}\in
G^{n}$, one can find a connection $A$ such that%
\[
\forall i=1,...,n:H(\alpha _{i},A)=g_{i}. 
\]
\end{remark}

For $\alpha \in \mathcal{L}_{\ast }$, let $T_{\alpha }:\mathcal{A}/\mathcal{G%
}\rightarrow \mathbb{C}$ be the Wilson loop function%
\[
T_{\alpha }(A):=\frac{1}{N}tr(\rho (H(\alpha ,A))), 
\]%
where $\rho $ is some matrix representation by $(N\times N)$ matrices of the
structure group $G$ of the principal bundle $P(M,G)$. Subsequently, $\rho $
will be supressed in the notation. Let $\mathcal{HA}$ be the algebra that is
generated by the $T_{\alpha }$. $\mathcal{HA}$ is called the \emph{holonomy
algebra}. Due to the fact that $G\subseteq U(N)$, we have%
\[
T_{\alpha }^{\ast }=T_{\alpha ^{-1}}, 
\]%
where $T_{\alpha }^{\ast }$ is defined by%
\[
T_{\alpha }^{\ast }(A):=\overline{T_{\alpha }(A)}. 
\]%
Thus $\mathcal{HA}$ is an involutive subalgebra of the $C^{\ast }$-algebra $%
\mathcal{B}(\mathcal{A}/\mathcal{G})$ of bounded complex-valued functions on 
$\mathcal{A}/\mathcal{G}$, equipped with the supremum norm.

\begin{theorem}
Each $\varphi \in Hom(\mathcal{L}_{\ast },G)$ induces a continuous algebra
homomorphism $\tau _{\varphi }:\mathcal{HA}\rightarrow \mathbb{C}$.
\end{theorem}

\begin{proof}
We show that the map%
\[
\tau _{\varphi }:T_{\alpha }\longmapsto \frac{1}{N}tr\ \varphi (\alpha )
\]%
fulfills the conditions (1) and (2) of Lemma 8 on the system of generators 
$\{T_{\alpha }|\alpha \in \mathcal{L}_{\ast }\}$. Let $\alpha
_{1},...,\alpha _{m},\beta _{1},...,\beta _{n}$ be hoops with $T_{\alpha
_{1}}...T_{\alpha _{m}}=T_{\beta _{1}}...T_{\beta _{n}}$. According to the
interpolation theorem, there exists some $A_{0}\in \mathcal{A}$ such that%
\begin{eqnarray*}
H(\alpha _{i},A_{0}) &=&\varphi (\alpha _{i}), \\
H(\beta _{j},A_{0}) &=&\varphi (\beta _{j})
\end{eqnarray*}%
for $i=1,...,m,\ j=1,...,n$. Then%
\begin{eqnarray*}
\tau _{\varphi }(T_{\alpha _{1}})...\tau _{\varphi }(T_{\alpha _{m}}) &=&%
\frac{1}{N}tr\ \varphi (\alpha _{1})...\frac{1}{N}tr\ \varphi (\alpha _{m})
\\
&=&\frac{1}{N}tr\ H(\alpha _{1},A_{0})...\frac{1}{N}tr\ H(\alpha _{m},A_{0})
\\
&=&T_{\alpha _{1}}(A_{0})...T_{\alpha _{m}}(A_{0}) \\
&=&T_{\beta _{1}}(A_{0})...T_{\beta _{n}}(A_{0}) \\
&=&\frac{1}{N}tr\ \varphi (\beta _{1})...\frac{1}{N}tr\ \varphi (\beta _{n})
\\
&=&\tau _{\varphi }(T_{\beta _{1}})...\tau _{\varphi }(T_{\beta n}).
\end{eqnarray*}%
Let $\tsum_{k}c_{k}\varphi _{k}=0$ for some $c_{k}\in \mathbb{C}$ and $%
p_{k}=T_{\alpha _{k_{1}}}...T_{\alpha _{k_{n_{k}}}}$. The interpolation
theorem assures the existence of some $A_{0}\in \mathcal{A}$ such that%
\[
H(\alpha _{k_{j}},A_{0})=\varphi (\alpha _{k_{j}})
\]%
for all $j\leq n_{k}$ and all $k$. Then%
\begin{eqnarray*}
\sum_{k}c_{k}\tau _{\varphi }(p_{k}) &=&\sum_{k}c_{k}\tau _{\varphi
}(T_{\alpha _{k_{1}}})...\tau _{\varphi }(T_{\alpha _{k_{n_{k}}}}) \\
&=&\sum_{k}c_{k}\frac{1}{N}tr\ \varphi (\alpha _{k_{1}})...\frac{1}{N}tr\
\varphi (\alpha _{k_{n_{k}}}) \\
&=&\sum_{k}c_{k}\frac{1}{N}tr\ H(\alpha _{k_{1}},A_{0})...\frac{1}{N}tr\
H(\alpha _{k_{n_{k}}},A_{0}) \\
&=&\sum_{k}c_{k}(T_{\alpha _{k_{1}}}...T_{\alpha _{k_{n_{k}}}})(A_{0}) \\
&=&(\sum_{k}c_{k}p_{k})(A_{0})=0.
\end{eqnarray*}%
For simplicity, the algebra homomorphism $\mathcal{HA}\rightarrow \mathbb{C}$
induced by $\tau _{\varphi }$ is denoted by $\tau _{\varphi }$, too. From
the considerations above, the continuity of $\tau _{\varphi }$ can be
inferred as follows:%
\begin{eqnarray*}
|\tau _{\varphi }(\sum_{k}c_{k}p_{k})| &=&|(\sum_{k}c_{k}p_{k})(A_{0})| \\
&\leq &\sup_{A\in \mathcal{A}/\mathcal{G}}|(\sum_{k}c_{k}p_{k})(A)| \\
&=&|(\sum_{k}c_{k}p_{k})|_{\infty }.
\end{eqnarray*}%
\newline
\end{proof}\newline

Let $C^{\ast }(\mathcal{HA})$ be the closure of $\mathcal{HA}$ in $\mathcal{B%
}(\mathcal{A}/\mathcal{G})$. $C^{\ast }(\mathcal{HA})$ is called the \emph{%
Ashtekar-Isham-Algebra }\cite{AshIsh92}. Each $\varphi \in Hom(\mathcal{L}%
_{\ast },G)$ gives a character of the commutative $C^{\ast }$-algebra $%
C^{\ast }(\mathcal{HA})$, denoted by $\tau _{\varphi }$ as above. We want to
examine how much the map%
\[
Hom(\mathcal{L}_{\ast },G)\rightarrow Spec(C^{\ast }(\mathcal{HA}))=:%
\overline{\mathcal{A}/\mathcal{G}} 
\]%
fails to be injective. For this, we need the compact group $\mathcal{M}(%
\mathcal{L}_{\ast })$ associated to the hoop group $\mathcal{L}_{\ast }$ as
described in section 2.\newline

First we will show that in our special case $H=\mathcal{L}_{\ast }$ the map $%
\iota :\mathcal{L}_{\ast }\rightarrow \mathcal{M}(\mathcal{L}_{\ast })$
defined in Prop. 5\ is injective. It is well known that the kernel of $\iota 
$ is equal to the intersection of the kernels of the continuous morphisms of 
$\mathcal{L}_{\ast }$ into all compact groups, see ch. 16.4 of \cite{Dix77}.
For reasons that will become clear later, we will equip $\mathcal{L}_{\ast }$
with the discrete topology. Thus the holonomy mappings $H(\_,A):\mathcal{L}%
_{\ast }\rightarrow G$ are contained in the set of continuous morphisms of $%
\mathcal{L}_{\ast }$ into compact groups.\newline

Remark 3 shows that for every $g\in G$ and every $\alpha \in \mathcal{L}%
_{\ast },\alpha \neq x$, there exists a connection $A$ such that $H(\_,A)=g$%
. So the intersection of the kernels of the holonomy mappings is $\{x\}$,
and $\iota $ is injective.\newline

Now back to the question of how much the map%
\[
Hom(\mathcal{L}_{\ast },G)\rightarrow Spec(C^{\ast }(\mathcal{HA})), 
\]%
defined on generators by%
\[
\tau _{\varphi }:T_{\alpha }\longmapsto \frac{1}{N}tr\ \varphi (\alpha ), 
\]%
fails to be injective. Let $\varphi _{1},\varphi _{2}\in Hom(\mathcal{L}%
_{\ast },G)$ and suppose that $\tau _{\varphi _{1}}=\tau _{\varphi _{2}}$.
Then%
\[
\forall \alpha \in \mathcal{L}_{\ast }:tr(\varphi _{1}(a))=tr(\varphi
_{2}(\alpha )), 
\]%
i.e.%
\[
\forall \alpha \in \mathcal{L}_{\ast }:tr(\mathcal{M}(\varphi _{1})(\iota
(\alpha )))=tr(\mathcal{M}(\varphi _{2})(\iota (\alpha ))), 
\]%
and hence, because $\mathcal{M}(\varphi _{1}),\mathcal{M}(\varphi _{2})$ are
continuous and $\iota (\mathcal{L}_{\ast })$ is dense in $\mathcal{M}(%
\mathcal{L}_{\ast })$:%
\[
\forall \xi \in \mathcal{M}(\mathcal{L}_{\ast }):tr(\mathcal{M}(\varphi
_{1})(\xi ))=tr(\mathcal{M}(\varphi _{2})(\xi )). 
\]%
Thus, the characters of the continuous unitary representations $\mathcal{M}%
(\varphi _{1}),\mathcal{M}(\varphi _{2}):\mathcal{M}(\mathcal{L}_{\ast
})\rightarrow U(N)$ of the compact group $\mathcal{M}(\mathcal{L}_{\ast })$
are the same and the representations are unitarily equivalent:%
\[
\exists T\in U(N)\ \forall \xi \in \mathcal{M}(\mathcal{L}_{\ast }):\mathcal{%
M}(\varphi _{2})(\xi )=T\mathcal{M}(\varphi _{1})(\xi )T^{\ast }. 
\]%
This is equivalent to%
\[
\forall \alpha \in \mathcal{L}_{\ast }:\varphi _{2}(\alpha )=T\varphi
_{1}(\alpha )T^{\ast } 
\]%
or $\varphi _{2}=T\varphi _{1}T^{\ast }$, for short. The homomorphisms $%
\varphi _{1},\varphi _{2}$ are called unitarily equivalent, too. If $\varphi
_{1},\varphi _{2}$ are unitarily equivalent in this way, one has $\tau
_{\varphi _{1}}=\tau _{\varphi _{2}}$, because%
\begin{eqnarray*}
\tau _{\varphi _{2}}(T_{\alpha }) &=&\frac{1}{N}tr\ \varphi _{2}(\alpha ) \\
&=&\frac{1}{N}tr(T\varphi _{1}(\alpha )T^{\ast }) \\
&=&\frac{1}{N}tr\ \varphi _{1}(\alpha ) \\
&=&\tau _{\varphi _{1}}(T_{\alpha })
\end{eqnarray*}%
for all $\alpha \in \mathcal{L}_{\ast }$. Unitary equivalence defines an
equivalence relation on $Hom(\mathcal{L}_{\ast },G)$, which is - by the
usual abuse of notation - denoted by $Ad$. The map $\varphi \mapsto \tau
_{\varphi }$ induces an injective map%
\[
\overline{\tau }:Hom(\mathcal{L}_{\ast },G)/Ad\rightarrow Spec(C^{\ast }(%
\mathcal{HA})). 
\]

\begin{lemma}
$Hom(\mathcal{L}_{\ast },G)/Ad$, equipped with the quotient topology defined
by the canonical projection%
\[
\pi :Hom(\mathcal{L}_{\ast },G)\rightarrow Hom(\mathcal{L}_{\ast },G)/Ad, 
\]%
is a compact space.
\end{lemma}

\begin{proof}
Since $Hom(\mathcal{L}_{\ast },G)$ is a compact space (see Theorem 7), it
suffices to show that $Hom(\mathcal{L}_{\ast },G)/Ad$, equipped with the
quotient topology, is a Hausdorff space. The group $U(N)$ acts on $Hom(%
\mathcal{L}_{\ast },U(N))$ by conjugation. Since $U(N)$ is compact, $Hom(%
\mathcal{L}_{\ast },U(N))/U(N)$ with the quotient topology is Hausdorff. The
imbedding%
\[
j:Hom(\mathcal{L}_{\ast },G)\rightarrow Hom(\mathcal{L}_{\ast },U(N))
\]%
obviously is continuous ($G$ is some closed subgroup of $U(N)$ by
definition) and induces an imbedding%
\[
\overline{j}:Hom(\mathcal{L}_{\ast },G)/Ad\rightarrow Hom(\mathcal{L}_{\ast
},U(N))/U(N)
\]%
by the commutative diagram\newline

\begin{center}
\setsqparms[1`1`1`1;1300`700]
\square[Hom(\QTR{cal}{L}_{\ast },G)`Hom(\QTR{cal}{L}_{\ast },U(N))`Hom(\QTR{cal}{L}_{\ast },G)/Ad`Hom(\QTR{cal}{L}_{\ast },U(N))/U(N);j`\pi`\pi _{U(N)}`\overline{j}]
\end{center}
%

The continuity of $\overline{j}$ follows from the universal property of the
quotient topology. As a direct consequence, $Hom(\mathcal{L}_{\ast },G)/Ad$
is a Hausdorff space.
\end{proof}

\begin{lemma}
Let $\mathcal{C}$ be a commutative $C^{\ast }$-algebra with $1$, $\mathcal{E}
$ a system of ge-nerators of $\mathcal{C}$, i.e. $lin\tprod\nolimits_{%
\mathcal{E}}\subseteq \mathcal{C}$ is dense in $\mathcal{C}$. Then the
Gelfand topology is the coarsest topology for which all $\hat{e}$ ($e\in 
\mathcal{E}$) are continuous.
\end{lemma}

\begin{proof}
The Gelfand topology on $Spec(\mathcal{C})$ is the coarsest topology for
which all Gelfand transforms $\hat{a}:Spec(\mathcal{C})\rightarrow \mathbb{C}
$ ($a\in \mathcal{C}$) are continuous. Hence the weak topology on $Spec(%
\mathcal{C})$ induced by the $\hat{e}$ ($e\in \mathcal{E}$), which is called
the $\mathcal{E}$-topology, is coarser than the Gelfand topology.\newline

On the other hand, the $\mathcal{E}$-topology is Hausdorff: $\sigma
_{1},\sigma _{2}\in Spec(\mathcal{C})$ and 
\begin{eqnarray*}
&\forall e\in &\mathcal{E}:\hat{e}(\sigma _{1})=\hat{e}(\sigma _{2}) \\
\Leftrightarrow  &\forall e\in &\mathcal{E}:\sigma _{1}(e)=\sigma _{2}(e)
\end{eqnarray*}%
implies $\sigma _{1}=\sigma _{2}.$ The functions $\hat{e}$ ($e\in \mathcal{E}
$) separate the points of $Spec(\mathcal{C})$, and thus the $\mathcal{E}$%
-topology is Hausdorff. Since the Gelfand topology is compact, and since to
a Hausdorff topology there is no strictly finer compact topology, the $%
\mathcal{E}$-topology and the Gelfand topology are identical. 
\end{proof}

\begin{lemma}
The map%
\[
\overline{\tau }:Hom(\mathcal{L}_{\ast },G)/Ad\rightarrow Spec(C^{\ast }(%
\mathcal{HA})) 
\]%
is continuous.
\end{lemma}

\begin{proof}
$\overline{\tau }$ is continuous if and only if $\tau =\overline{\tau }\circ
\pi $ is continuous. Since the Wilson loop functions $T_{\alpha }$ ($\alpha
\in \mathcal{L}_{\ast }$) generate the holonomy $C^{\ast }$-algebra $C^{\ast
}(\mathcal{HA})$, the Gelfand topology on $Spec(C^{\ast }(\mathcal{HA}))$ is
the coarsest topology for which all functions%
\[
\hat{T}_{\alpha }:Spec(C^{\ast }(\mathcal{HA}))\rightarrow \mathbb{C},\quad
\sigma \mapsto \sigma (T_{\alpha })
\]%
are continuous. Thus the map $\tau :Hom(\mathcal{L}_{\ast },G)\rightarrow
Spec(C^{\ast }(\mathcal{HA}))$ is continuous if and only if all functions%
\[
\hat{T}_{\alpha }\circ \tau :Hom(\mathcal{L}_{\ast },G)\rightarrow \mathbb{C}
\]%
are continuous. Now%
\[
(\hat{T}_{\alpha }\circ \tau )(\varphi )=\tau _{\varphi }(T_{\alpha })=\frac{%
1}{N}tr(\varphi (\alpha )).
\]%
This simply is the composition of the projection%
\[
pr_{\alpha }:Hom(\mathcal{L}_{\ast },G)\rightarrow G,\quad \varphi \mapsto
\varphi (\alpha )
\]%
and the normed trace map%
\[
G\rightarrow \mathbb{C},\quad g\mapsto \frac{1}{N}tr\ g,
\]%
and these maps are continuous.
\end{proof}

\bigskip Summing up, we have

\begin{theorem}
The quotient $Hom(\mathcal{L}_{\ast },G)/Ad$ is compact and continuously
embedded in $Spec(C^{\ast }(\mathcal{HA}))$ via the map%
\[
\overline{\tau }:\varphi \func{mod}Ad\mapsto \tau _{\varphi }. 
\]%
$\blacksquare $
\end{theorem}

We now show how $\mathcal{A}/\mathcal{G}$ can be regarded as a part of $Hom(%
\mathcal{L}_{\ast },G)/Ad$: the holonomy provides a canonical map%
\[
\mathcal{A}\rightarrow Hom(\mathcal{L}_{\ast },G),\quad A\mapsto H(\_,A). 
\]%
Assume that $A_{1}$ is gauge equivalent to $A_{2}.$ Then%
\[
A_{2}=\phi ^{\ast }A_{1} 
\]%
for some gauge transformation $\phi :P\rightarrow G$. For the holonomy map
of \ hoops based at $\ast $ one gets%
\[
H(\_,A_{2})=a_{\phi }(p_{0})^{-1}H(\_,A_{1})a_{\phi }(p_{0}), 
\]%
where $p_{0}\in P$ is a freely chosen but fixed point of the fiber over $%
\ast $, serving as initial value for the horizontal lifts of hoops in $M$.
Gauge equivalent connections give unitarily equivalent homomorphisms $%
\mathcal{L}_{\ast }\rightarrow G$. We have found inclusions%
\[
\overline{\tau }(\mathcal{A}/\mathcal{G})\subseteq \overline{\tau }(Hom(%
\mathcal{L}_{\ast },G)/Ad)\subseteq Spec(C^{\ast }(\mathcal{HA})). 
\]%
(One has%
\begin{eqnarray*}
\overline{\tau }(A\func{mod}\mathcal{G})(T_{\alpha }) &=&\tau
_{H(\_,A)}(T_{\alpha }) \\
&=&\frac{1}{N}tr\ H(\alpha ,A) \\
&=&T_{\alpha }(A),
\end{eqnarray*}%
so $\overline{\tau }(A\func{mod}\mathcal{G})=\varepsilon _{A\func{mod}%
\mathcal{G}}$ is the evaluation funtional at $A\func{mod}\mathcal{G}$.)
Accor-ding to a result of Rendall \cite{Ren93}, $\overline{\tau }(\mathcal{A}%
/\mathcal{G})$ is dense in $Spec(C^{\ast }(\mathcal{HA}))$, the space $%
\overline{\tau }(Hom(\mathcal{L}_{\ast },G)/Ad)$ is compact according to the
last theorem and hence closed in $Spec(C^{\ast }(\mathcal{HA}))$. Thus we get%
\begin{eqnarray*}
Spec(C^{\ast }(\mathcal{HA})) &=&\overline{\overline{\tau }(\mathcal{A}/%
\mathcal{G)}} \\
&\subseteq &\overline{\overline{\tau }(Hom(\mathcal{L}_{\ast },G)/Ad)} \\
&=&\overline{\tau }(Hom(\mathcal{L}_{\ast },G)/Ad),
\end{eqnarray*}%
i.e.%
\[
\overline{\tau }(Hom(\mathcal{L}_{\ast },G)/Ad)=Spec(C^{\ast }(\mathcal{HA}%
)). 
\]%
This proves

\begin{theorem}
The spectrum of the holonomy $C^{\ast }$-algebra $C^{\ast }(\mathcal{HA})$
is homoemorphic to the compact space $Hom(\mathcal{L}_{\ast },G)/Ad$.\newline
\end{theorem}

This result is not new, it was already found in \cite{AshLew94}. It is
central to all further developments in loop quantum gravity \cite{ALMMT95}.
In contrast to the known proofs, we have to refer neither to the Mandelstam
identities, \ nor the results of Giles \cite{Gil81} nor projective
techniques as first used in \cite{AshLew95}.\newline

Since Theorem 15 states a homeomorphism between $Spec(C^{\ast }(\mathcal{HA}%
))$ and the set $Hom(\mathcal{L}_{\ast },G)/Ad$ of \emph{all }equivalence
classes of homomorphisms from the hoop group $\mathcal{L}_{\ast }$ to $G$,
it is natural to equip $\mathcal{L}_{\ast }$ with the discrete topology.%
\newline

We have seen in Lemma 6 that there is a canonical bijection%
\[
Hom_{c}(\mathcal{M}(\mathcal{L}_{\ast }),G)\simeq Hom(\mathcal{L}_{\ast
},G), 
\]%
which of course induces a canonical bijection%
\[
Hom_{c}(\mathcal{M}(\mathcal{L}_{\ast }),G)/Ad\simeq Hom(\mathcal{L}_{\ast
},G)/Ad. 
\]%
The space $Hom(\mathcal{L}_{\ast },G)$ is equipped with the relative
topology induced by the product topology on $G^{\mathcal{L}_{\ast }}$ (see
Theorem 7), and is a compact space in this topology. $Hom(\mathcal{L}_{\ast
},G)/Ad$ is compact in the quotient topology (Lemma 11). $Hom_{c}(\mathcal{M}%
(\mathcal{L}_{\ast }),G)/Ad$ canonically is equipped with the topology
induced by the above bijection - which of course is a homoemorphism, then -,
and thus becomes a compact space, too. So we have a homeomorphism%
\[
\overline{\mathcal{A}/\mathcal{G}}=Spec(C^{\ast }(\mathcal{HA))}\simeq
Hom_{c}(\mathcal{M}(\mathcal{L}_{\ast }),G)/Ad. 
\]

\section{Generalized Wilson loop functions and the Hilbert algebra $L_{2}(%
\mathcal{M}(\mathcal{L}_{\ast }))$}

In this section, we will introduce the notion of \textit{generalized Wilson
loop functions}, which are characters of representations $\rho _{h}:\mathcal{%
M}(\mathcal{L}_{\ast })\rightarrow SU(2)$. The ordinary Wilson loop
functions simply are characters of representations $\rho :\mathcal{L}_{\ast
}\rightarrow G$. We specialize to the case $G=SU(2)$ now. This is the gauge
group needed in loop quantum gravity \cite{Ash86,Ash87}.\newline

The elements of $\overline{\mathcal{A}/\mathcal{G}}=Spec(C^{\ast }(\mathcal{%
HA}))$ can be understood as (generalized gauge equivalence classes of)
generalized connections \cite{AshIsh92}. We have seen in section 3 that to
every $h\in \overline{\mathcal{A}/\mathcal{G}}$ there exists some unique
function $\frac{1}{2}tr\circ \rho $, where $\rho $ is a homomorphism from $%
\mathcal{L}_{\ast }$ to $SU(2)$. Taking the trace of course corresponds to
the $\func{mod}Ad$ operation, since every equivalence class of
representations is umambiguously characterized by its trace. On $\mathcal{HA}
$ one has%
\[
\forall \alpha \in \mathcal{L}_{\ast }:h(T_{\alpha })=\frac{1}{2}tr(\rho
(\alpha )). 
\]

Since we have found $Hom(\mathcal{L}_{\ast },SU(2))/Ad\cong Hom_{c}(\mathcal{%
M}(\mathcal{L}_{\ast }),SU(2))/Ad$, this means that the elements $h\in 
\overline{\mathcal{A}/\mathcal{G}}$ correspond bijectively to the functions%
\[
\frac{1}{2}tr\circ \rho _{h}, 
\]%
where%
\[
\rho _{h}:\mathcal{M}(\mathcal{L}_{\ast })\rightarrow SU(2) 
\]%
is a continuous representation of the compact group $\mathcal{M}:=\mathcal{M}%
(\mathcal{L}_{\ast })$. The function 
\[
t_{h}:=tr\circ \rho _{h} 
\]%
is the character of the representation $\rho _{h}$ of $\mathcal{M}$.

\begin{definition}
Let $\rho :\mathcal{M}\rightarrow SU(2)$ be a continous representation. Then%
\[
t_{\rho }:=tr\circ \rho :\mathcal{M}\longrightarrow \mathbb{R} 
\]%
is called a \textbf{generalized Wilson loop function}.
\end{definition}

Note that in contrast to some of the existing literature on Wilson loop
functions, we do not include a numerical factor $\frac{1}{2}$ in this
definition. Thus the trace map is not normed. For our case, $G=SU(2)$, this
means that $t_{\rho }$ is bounded between $-2$ and $2$, not $-1$ and $1$.%
\newline

But there is a much more important change: instead of regarding Wilson loop
functions as functions of (gauge equivalence classes of) connections, with
hoops serving as an index, we regard functions of hoops (or elements of $%
\mathcal{M}=\mathcal{M}(\mathcal{L}_{\ast })$, to be more precise). Here the
connection, given by the corresponding element $\rho $ of $Hom_{c}(\mathcal{M%
}(\mathcal{L}_{\ast }),G)$, plays the role of an index. A more suggestive
notation is%
\[
t_{A}:\mathcal{M}\longrightarrow \mathbb{R}, 
\]%
where $A\in \mathcal{A}/\mathcal{G}$, in contrast to%
\[
T_{\alpha }:\mathcal{A}/\mathcal{G}\longrightarrow \mathbb{R} 
\]%
(for other groups than $SU(2)$ the mappings $t_{A}$ and $T_{\alpha }$ are
complex-valued, of course).\newline

Let $d\xi $ be the normed Haar measure on $\mathcal{M}$. $L_{2}(\mathcal{M}%
):=L_{2}(\mathcal{M},d\xi )$ is the Hilbert algebra of $\mathcal{M}$, where
multiplication is given by convolution:%
\begin{eqnarray*}
(f\star g)(\xi ) &=&\int_{\mathcal{M}}f(\xi \eta ^{-1})g(\eta )d\eta \\
&=&\int_{\mathcal{M}}f(\xi \eta )g(\eta ^{-1})d\eta .
\end{eqnarray*}%
Since $tr(AB)=tr(BA)$, one has%
\[
\forall \xi ,\eta \in \mathcal{M}:t_{\rho }(\xi \eta )=t_{\rho }(\eta \xi ), 
\]%
so the generalized Wilson loop functions belong to the center of the algebra 
$L_{2}(\mathcal{M})$. This holds for more general groups than $SU(2)$, but
the following result is specific to two-dimensional representations, since
it uses some trace identities for complex $2\times 2$-matrices of
determinant $1$:%
\begin{eqnarray*}
trC\ trD &=&trCD+trCD^{-1}, \\
trC &=&trC^{-1}.
\end{eqnarray*}
We bother to write down the next theorem and some corollaries explicitly,
although results of this kind may be found in books like \cite{HewRos70}.
The proof given here is adapted to our situation.

\begin{theorem}
Let $f\in L_{2}(\mathcal{M})$ be a symmetric function, i.e. $f(\xi )=f(\xi
^{-1})$. Then%
\[
2t_{\rho }\star f=\left\langle f,t_{\rho }\right\rangle t_{\rho } 
\]%
for all generalized Wilson loop functions $t_{\rho }$ on $\mathcal{M}$.
\end{theorem}

\begin{proof}
One has%
\begin{eqnarray*}
(t_{\rho }\star f)(\xi ) &=&\int tr\ \rho (\xi \eta ^{-1})f(\eta )d\eta  \\
&=&\int tr(\rho (\xi )\rho (\eta ^{-1}))f(\eta )d\eta  \\
&=&\int tr\ \rho (\xi )\ tr\ (\rho (\eta )^{-1})f(\eta )d\eta -\int tr(\rho
(\xi )\rho (\eta ))f(\eta )d\eta  \\
&=&tr\ \rho (\xi )\int f(\eta )tr\ \rho (\eta )d\eta -\int tr\ \rho (\xi
\eta )f(\eta ^{-1})d\eta  \\
&=&tr\ \rho (\xi )\left\langle f,tr\circ \rho \right\rangle -(t_{\rho }\star
f)(\xi ) \\
&=&t_{\rho }(\xi )\left\langle f,t_{\rho }\right\rangle -(t_{\rho }\star
f)(\xi ),
\end{eqnarray*}%
and hence%
\[
2t_{\rho }\star f=\left\langle f,t_{\rho }\right\rangle t_{\rho }.
\]%
In the fourth step above, we used the symmetry of $f$ and $t_{\rho }$:%
\begin{eqnarray*}
t_{\rho }(\xi ) &=&tr\ \rho (\xi ) \\
&=&tr\ \rho (\xi )^{-1} \\
&=&tr\ \rho (\xi ^{-1}) \\
&=&t_{\rho }(\xi ^{-1}).
\end{eqnarray*}
\end{proof}

\begin{remark}
The last result can be generalized to all $f\in L_{2}(\mathcal{M})$ in the
following sense:
\end{remark}

For $f\in L_{2}(\mathcal{M)}$, let $f^{\ast }\in L_{2}(\mathcal{M})$ be
defined by%
\[
f^{\ast }(\eta ):=\overline{f(\eta ^{-1})}. 
\]%
Then%
\begin{eqnarray*}
&&\int t_{\rho }(\xi \eta )f(\eta )d\eta \\
&=&\int t_{\rho }(\xi \eta )\overline{f^{\ast }(\eta ^{-1})}d\eta \\
&=&(t_{\rho }\star \overline{f^{\ast }})(\xi )
\end{eqnarray*}%
and one obtains the more general formula%
\[
t_{\rho }\star (f+\overline{f^{\ast }})=\left\langle f,t_{\rho
}\right\rangle t_{\rho }. 
\]%
Writing $\check{f}(\eta ):=f(\eta ^{-1})$, this gives%
\[
t_{\rho }\star (f+\check{f})=\left\langle f,t_{\rho }\right\rangle t_{\rho
}, 
\]%
and%
\[
2t_{\rho }\star t_{\sigma }=\left\langle t_{\sigma },t_{\rho }\right\rangle
t_{\rho }. 
\]

\begin{corollary}
For all continuous representations $\rho ,\sigma :\mathcal{M}\rightarrow
SU(2)$, one has%
\[
2t_{\rho }\star t_{\sigma }=\left\langle t_{\sigma },t_{\rho }\right\rangle
t_{\rho }. 
\]%
If $t_{\rho }\neq t_{\sigma }$, i.e. if the representations $\rho $ and $%
\sigma $ are not equivalent, then%
\[
\left\langle t_{\sigma },t_{\rho }\right\rangle =0. 
\]
\end{corollary}

\begin{proof}
The first equality was shown above. The orthogonality results from the
commutativity of the convolution:%
\begin{eqnarray*}
\left\langle t_{\sigma },t_{\rho }\right\rangle t_{\rho } &=&2t_{\rho }\star
t_{\sigma } \\
&=&2t_{\sigma }\star t_{\rho } \\
&=&\left\langle t_{\rho },t_{\sigma }\right\rangle t_{\sigma } \\
&=&\left\langle t_{\sigma },t_{\rho }\right\rangle t_{\sigma }
\end{eqnarray*}%
and hence $\left\langle t_{\sigma },t_{\rho }\right\rangle =0$ for $%
t_{\sigma }\neq t_{\rho }$.
\end{proof}\newline

When $t_{\rho }$ is normed by passing to%
\[
e_{\rho }:=\frac{2}{\left\langle t_{\rho },t_{\rho }\right\rangle }t_{\rho
}, 
\]%
one gets

\begin{corollary}
The elements $e_{\rho }$ of $L_{2}(\mathcal{M})$ are idempotent with respect
to the convolution:%
\[
e_{\rho }\star e_{\rho }=e_{\rho }. 
\]%
If $\rho $ and $\sigma $ are non-equivalent representations, then $e_{\rho }$
and $e_{\sigma }$ are orthogonal,%
\[
\left\langle e_{\rho },e_{\sigma }\right\rangle =0. 
\]
\end{corollary}

\begin{proof}
One has%
\begin{eqnarray*}
e_{\rho }\star e_{\rho } &=&\frac{4}{\left\langle t_{\rho },t_{\rho
}\right\rangle ^{2}}t_{\rho }\star t_{\rho } \\
&=&\frac{2}{\left\langle t_{\rho },t_{\rho }\right\rangle ^{2}}2t_{\rho
}\star t_{\rho } \\
&=&\frac{2}{\left\langle t_{\rho },t_{\rho }\right\rangle ^{2}}\left\langle
t_{\rho },t_{\rho }\right\rangle t_{\rho } \\
&=&e_{\rho }.
\end{eqnarray*}
\end{proof}\newline

The generalized Wilson loop functions $t_{\rho }$ are easily seen to be
continuous functions $\mathcal{M}(\mathcal{L}_{\ast })\rightarrow \mathbb{C}$%
, so they come from almost periodic functions on $\mathcal{L}_{\ast }$:
consider the commutative diagram

\begin{center}
\settripairparms[1`1`1`1`-1;900]
\Vtrianglepair[\QTR{cal}{L}_{\ast }`\QTR{cal}{M}(\QTR{cal}{L}_{\ast })`\QTR{Bbb}{C}`G;\iota`tr\circ \QTR{cal}{M}(\varphi )`\varphi `\QTR{cal}{M}(\varphi )`tr]
\end{center}%

Here $\varphi \in Hom(\mathcal{L}_{\ast },G)$. Lemma 6 shows the existence
of a continuous homomorphism $\mathcal{M}(\varphi ):\mathcal{M}(\mathcal{L}%
_{\ast })\rightarrow G$. The trace function $tr$ is continuous, and thus $%
tr\circ \mathcal{M}(\varphi ):\mathcal{M}(\mathcal{L}_{\ast })\rightarrow 
\mathbb{C}$ is continuous. It follows from Prop. 5 that the function $%
tr\circ \varphi :\mathcal{L}_{\ast }\rightarrow \mathbb{C}$ is almost
periodic and the function $t_{\rho }=t_{\mathcal{M}(\varphi )}=tr\circ 
\mathcal{M}(\varphi )$ is its extension to $\mathcal{M}(\mathcal{L}_{\ast })$%
.\newline

It seems quite suggestive to regard $L_{2}(\mathcal{M})$ as a candidate for
the Hilbert space of loop quantum gravity in the loop representation. To
show that this is the right choice, several issues have to be clarified. The
most important is the loop transform \cite{RovSmo90,AshLew94} that
translates from the connection representation to the loop representation.
The loop transform, an appropriate extension of it and the inverse loop
transform will be treated in a subsequent paper. Another important issue is
diffeomorphism invariance. Since the ultimate goal is some form of quantum
gravity, one has to regard the symmetries of the classical theory, which is
General Relativity in a Hamiltonian formulation. In approaches of this kind,
spacetime $M$ is assumed to be globally hyperbolic, so that according to a
theorem by Geroch \cite{HawEll73} one has%
\[
M\simeq \mathbb{R}\times \Sigma , 
\]%
where $\Sigma $ is a compact three-dimensional Riemannian manifold. The
above isomorphism is not given canonically, but is the foliation of
spacetime into space $\Sigma $ and time $\mathbb{R}$ as seen by some
observer. Another observer corresponds to another foliation. The theory is
formulated with respect to some fixed foliation. For example, the
connections regarded throughout this paper are connections on a principal
bundle over $\Sigma $. Physical predictions of the theory must be
independent of the choice of foliation, of course. Since we are concerned
with the kinematical aspects of the theory, we do not have to consider the
full diffeomorphism group of spacetime, but merely the diffeomorphisms%
\[
\phi :\Sigma \longrightarrow \Sigma 
\]%
of space. Since we have regarded piecewise analytic hoops throughout, we
will only consider analytic diffeomorphisms. If $L_{2}(\mathcal{M})$ shall
serve as Hilbert space for the loop representation, it is a necessary
condition that its Haar measure is diffeomorphism invariant:

\begin{proposition}
The Haar measure $\xi $ on $\mathcal{M}(\mathcal{L}_{\ast })$ is invariant
under analytic diffeomorphisms $\phi :\Sigma \rightarrow \Sigma $.
\end{proposition}

\begin{proof}
Let $Diff_{an}(\Sigma )$ be the group of analytic diffeomorphisms of $\Sigma 
$. $Diff_{an}(\Sigma )$ operates by isomorphisms on $\mathcal{L}_{\ast }$:
let $\gamma \in \mathcal{L}_{\ast }$. Then%
\[
\tilde{\phi}:\mathcal{L}_{\ast }\rightarrow \mathcal{L}_{\ast },\quad \gamma
\mapsto (\phi \circ \gamma )
\]%
and $\tilde{\phi}(\gamma \delta )=(\tilde{\phi}(\gamma ))(\tilde{\phi}%
(\delta ))$. The following diagram is commutative:

\begin{center}
\setsqparms[1`1`1`1;1300`700]
\square[\QTR{cal}{L}_{\ast }`\QTR{cal}{L}_{\ast }`\QTR{cal}{M}(\QTR{cal}{L}_{\ast })`\QTR{cal}{M}(\QTR{cal}{L}_{\ast });\tilde{\phi}`\iota `\iota `\QTR{cal}{M}(\phi )]
\end{center}%

The mapping $\iota \circ \tilde{\phi}:\mathcal{L}_{\ast }\rightarrow 
\mathcal{M}(\mathcal{L}_{\ast })$ is a homomorphism from $\mathcal{L}_{\ast }
$ to $\mathcal{M}(\mathcal{L}_{\ast })$ and induces a continuous
homomorphism $\mathcal{M}(\phi ):\mathcal{M}(\mathcal{L}_{\ast })\rightarrow 
\mathcal{M}(\mathcal{L}_{\ast })$, which is an automorphism of $\mathcal{M}(%
\mathcal{L}_{\ast })$, of course. Hence $Diff_{an}(\Sigma )$ induces
continuous automorphisms of $\mathcal{M}(\mathcal{L}_{\ast })$. Let $\phi
,\psi \in Diff_{an}(\Sigma )$. Then one has $\mathcal{M}\mathbf{(}\phi \psi
)=\mathcal{M}(\phi )\mathcal{M}(\psi )$, which gives%
\begin{eqnarray*}
\mathcal{M}:Diff_{an}(\Sigma ) &\rightarrow &Aut_{c}(\mathcal{M}(\mathcal{L}%
_{\ast })), \\
\phi  &\mapsto &\mathcal{M}(\phi ).
\end{eqnarray*}%
As is well known, the Haar measure on a compact group is invariant under
such automorphisms of the group. Let $f\in C(\mathcal{M}(\mathcal{L}_{\ast
}))$. So we have%
\[
\int_{\mathcal{M}(\mathcal{L}_{\ast })}(f\circ \mathcal{M}(\phi ))(\xi )d\xi
=\int_{\mathcal{M}(\mathcal{L}_{\ast })}f(\xi )d\xi .
\]
\end{proof}

\section{Outlook}

As already mentioned above, in a subsequent paper we will define an extended
loop transform from the Hilbert space $L_{2}(\overline{\mathcal{A}/\mathcal{G%
}},d\mu )$ of loop quantum gravity in the connection representation to the
Hilbert space $L_{2}(\mathcal{M})$. This will employ the extension of the
Wilson loop functions $T_{\alpha }$ to $\mathcal{M}(\mathcal{L}_{\ast })$.
(Here the range of the index is extended.) The inverse extended loop
transform is then given canonically, confirming the claim that $L_{2}(%
\mathcal{M})$ should be regarded as the Hilbert space of loop quantum
gravity in the loop representation.\newline

Furthermore, we will discuss some aspects of the implementation of the
diffeomorphism constraints. In \cite{AshLew94}, there are remarks on the
''dual'' r\^{o}les of the hoop group $\mathcal{L}_{\ast }$ and the quantum
configuration space $\overline{\mathcal{A}/\mathcal{G}}$. We will present
some results on the action of the diffeomorphism group making this point
clearer.

\section{Acknowledgements}

AD was partially supported by the Studienstiftung des Deutschen Volkes.

\end{document}